%%%%%%%%%%%%%%%%%%%%%%%%%%%%%%%%%%%%%%%%%%%%%%%%

%harvmac
%\input harvmac

%%%%%%%%%%%%%%%%%%  tex macros for preprints, cm version %%%%%%%%%%%%%%
%                     (P. Ginsparg, last updated 9/91)
%                if confused, type `b' in response to query
%
%---------------------------------------------------------------------%
%% site dependent options:
%% \unredoffs and \redoffs define horizontal and vertical offsets
%% respectively for unreduced and reduced modes. \speclscape defines
%% the \special{} call that sets printer to landscape (sideways) mode.
%% from standard set below, leave uncommented as appropriate or redefine
%
%%% next 400dpi
%\def\unredoffs{} \def\redoffs{\voffset=-.31truein\hoffset=-.48truein}
%\def\speclscape{\special{landscape}}
%
%%% apple lw
\def\unredoffs{} \def\redoffs{\voffset=-.31truein\hoffset=-.59truein}
\def\speclscape{\special{ps: landscape}}
%
%%% qms lasergrafix:
%\def\unredoffs{} \def\redoffs{\voffset=-.4truein\hoffset=.125truein}
%\def\speclscape{\special{qms: landscape}}
%
%%% saclay A4 paper:
%\def\unredoffs{\hoffset-.14truein\voffset-.2truein}
%\def\redoffs{\voffset=-.45truein\hoffset=-.21truein}
%\def\speclscape{\special{landscape}}
%
%---------------------------------------------------------------------%
%
\newbox\leftpage \newdimen\fullhsize \newdimen\hstitle \newdimen\hsbody
\tolerance=1000\hfuzz=2pt
\catcode`\@=11 % This allows us to modify PLAIN macros.
\def\bigans{b }
\def\answ{b }

%\message{ big or little (b/l)? }\read-1 to\answ
%
\ifx\answ\bigans\message{(This will come out unreduced.}
\magnification=1200\unredoffs\baselineskip=16pt plus 2pt minus 1pt
\hsbody=\hsize \hstitle=\hsize %take default values for unreduced format
\else\message{(This will be reduced.} \let\l@r=L
\magnification=1000\baselineskip=16pt plus 2pt minus 1pt \vsize=7truein
\redoffs \hstitle=8truein\hsbody=4.75truein\fullhsize=10truein\hsize=\hsbody
\output={\ifnum\pageno=0 %%% This is the HUTP version
  \shipout\vbox{\speclscape{\hsize\fullhsize\makeheadline}
    \hbox to \fullhsize{\hfill\pagebody\hfill}}\advancepageno
  \else
  \almostshipout{\leftline{\vbox{\pagebody\makefootline}}}\advancepageno
  \fi}
\def\almostshipout#1{\if L\l@r \count1=1 \message{[\the\count0.\the\count1]}
      \global\setbox\leftpage=#1 \global\let\l@r=R
 \else \count1=2
  \shipout\vbox{\speclscape{\hsize\fullhsize\makeheadline}
      \hbox to\fullhsize{\box\leftpage\hfil#1}}  \global\let\l@r=L\fi}
\fi
%---------------------------------------------------------------------
%
\newcount\yearltd\yearltd=\year\advance\yearltd by -1900

\def\Title#1#2{\nopagenumbers\abstractfont\hsize=\hstitle\rightline{#1}%
\vskip 1in\centerline{\titlefont #2}\abstractfont\vskip .5in\pageno=0}
\def\Date#1{\vfill\leftline{#1}\tenpoint\supereject\global\hsize=\hsbody%
\footline={\hss\tenrm\folio\hss}}%  restores pagenumbers
%
%       use following instead of \Date on the preliminary draft,
%       puts date/time on each page in big mode, writes labels in margins

\def\draftmode{\message{ DRAFTMODE }\def\draftdate{{\rm preliminary draft:
\number\month/\number\day/\number\yearltd\ \ \hourmin}}%
\headline={\hfil\draftdate}\writelabels\baselineskip=20pt plus 2pt minus 2pt
 {\count255=\time\divide\count255 by 60 \xdef\hourmin{\number\count255}
  \multiply\count255 by-60\advance\count255 by\time
  \xdef\hourmin{\hourmin:\ifnum\count255<10 0\fi\the\count255}}}
%       use \nolabels to get rid of eqn, ref, and fig labels in draft mode
\def\nolabels{\def\wrlabeL##1{}\def\eqlabeL##1{}\def\reflabeL##1{}}
\def\writelabels{\def\wrlabeL##1{\leavevmode\vadjust{\rlap{\smash%
{\line{{\escapechar=` \hfill\rlap{\sevenrm\hskip.03in\string##1}}}}}}}%
\def\eqlabeL##1{{\escapechar-1\rlap{\sevenrm\hskip.05in\string##1}}}%
\def\reflabeL##1{\noexpand\llap{\noexpand\sevenrm\string\string\string##1}}}
\nolabels
%
% tagged sec numbers
\global\newcount\secno \global\secno=0
\global\newcount\meqno \global\meqno=1
\def\newsec#1{\global\advance\secno by1\message{(\the\secno. #1)}
%\ifx\answ\bigans \vfill\eject \else \bigbreak\bigskip \fi  %if desired
\global\subsecno=0\eqnres@t\noindent{\bf\the\secno. #1}
\writetoca{{\secsym} {#1}}\par\nobreak\medskip\nobreak}
\def\eqnres@t{\xdef\secsym{\the\secno.}\global\meqno=1\bigbreak\bigskip}
\def\sequentialequations{\def\eqnres@t{\bigbreak}}\xdef\secsym{}
\global\newcount\subsecno \global\subsecno=0
\def\subsec#1{\global\advance\subsecno by1\message{(\secsym\the\subsecno. #1)}
\ifnum\lastpenalty>9000\else\bigbreak\fi
\noindent{\it\secsym\the\subsecno. #1}\writetoca{\string\quad
{\secsym\the\subsecno.} {#1}}\par\nobreak\medskip\nobreak}
\def\appendix#1#2{\global\meqno=1\global\subsecno=0\xdef\secsym{\hbox{#1.}}
\bigbreak\bigskip\noindent{\bf Appendix #1. #2}\message{(#1. #2)}
\writetoca{Appendix {#1.} {#2}}\par\nobreak\medskip\nobreak}
%
%       \eqn\label{a+b=c}   gives displayed equation, numbered
%               consecutively within sections.
%     \eqnn and \eqna define labels in advance (of eqalign?)
%
\def\eqnn#1{\xdef #1{(\secsym\the\meqno)}\writedef{#1\leftbracket#1}%
\global\advance\meqno by1\wrlabeL#1}
\def\eqna#1{\xdef #1##1{\hbox{$(\secsym\the\meqno##1)$}}
\writedef{#1\numbersign1\leftbracket#1{\numbersign1}}%
\global\advance\meqno by1\wrlabeL{#1$\{\}$}}
\def\eqn#1#2{\xdef #1{(\secsym\the\meqno)}\writedef{#1\leftbracket#1}%
\global\advance\meqno by1$$#2\eqno#1\eqlabeL#1$$}
%
%            footnotes
\newskip\footskip\footskip14pt plus 1pt minus 1pt %sets footnote baselineskip
\def\footnotefont{\ninepoint}\def\f@t#1{\footnotefont #1\@foot}
\def\f@@t{\baselineskip\footskip\bgroup\footnotefont\aftergroup\@foot\let\next}
\setbox\strutbox=\hbox{\vrule height9.5pt depth4.5pt width0pt}
\global\newcount\ftno \global\ftno=0
\def\foot{\global\advance\ftno by1\footnote{$^{\the\ftno}$}}
%
%say \footend to put footnotes at end
%will cause problems if \ref used inside \foot, instead use \nref before
\newwrite\ftfile
\def\footend{\def\foot{\global\advance\ftno by1\chardef\wfile=\ftfile
$^{\the\ftno}$\ifnum\ftno=1\immediate\openout\ftfile=foots.tmp\fi%
\immediate\write\ftfile{\noexpand\smallskip%
\noexpand\item{f\the\ftno:\ }\pctsign}\findarg}%
\def\footatend{\vfill\eject\immediate\closeout\ftfile{\parindent=20pt
\centerline{\bf Footnotes}\nobreak\bigskip\input foots.tmp }}}
\def\footatend{}
%
%     \ref\label{text}
% generates a number, assigns it to \label, generates an entry.
% To list the refs on a separate page,  \listrefs
%
\global\newcount\refno \global\refno=1
\newwrite\rfile
\def\ref{[\the\refno]\nref}
\def\nref#1{\xdef#1{[\the\refno]}\writedef{#1\leftbracket#1}%
\ifnum\refno=1\immediate\openout\rfile=refs.tmp\fi
\global\advance\refno by1\chardef\wfile=\rfile\immediate
\write\rfile{\noexpand\item{#1\ }\reflabeL{#1\hskip.31in}\pctsign}\findarg}
%   horrible hack to sidestep tex \write limitation
\def\findarg#1#{\begingroup\obeylines\newlinechar=`\^^M\pass@rg}
{\obeylines\gdef\pass@rg#1{\writ@line\relax #1^^M\hbox{}^^M}%
\gdef\writ@line#1^^M{\expandafter\toks0\expandafter{\striprel@x #1}%
\edef\next{\the\toks0}\ifx\next\em@rk\let\next=\endgroup\else\ifx\next\empty%
\else\immediate\write\wfile{\the\toks0}\fi\let\next=\writ@line\fi\next\relax}}
\def\striprel@x#1{} \def\em@rk{\hbox{}}
\def\lref{\begingroup\obeylines\lr@f}
\def\lr@f#1#2{\gdef#1{\ref#1{#2}}\endgroup\unskip}

\def\addref#1{\immediate\write\rfile{\noexpand\item{}#1}} %now unnecessary
\def\startrefs#1{\immediate\openout\rfile=refs.tmp\refno=#1}
\def\xref{\expandafter\xr@f}\def\xr@f[#1]{#1}
\def\refs#1{\count255=1[\r@fs #1{\hbox{}}]}
\def\r@fs#1{\ifx\und@fined#1\message{reflabel \string#1 is undefined.}%
\nref#1{need to supply reference \string#1.}\fi%
\vphantom{\hphantom{#1}}\edef\next{#1}\ifx\next\em@rk\def\next{}%
\else\ifx\next#1\ifodd\count255\relax\xref#1\count255=0\fi%
\else#1\count255=1\fi\let\next=\r@fs\fi\next}
%

%
% this is ugly, but moore insists
\newwrite\ffile\global\newcount\figno \global\figno=1
\def\fig{fig.~\the\figno\nfig}
\def\nfig#1{\xdef#1{fig.~\the\figno}%
\writedef{#1\leftbracket fig.\noexpand~\the\figno}%
\ifnum\figno=1\immediate\openout\ffile=figs.tmp\fi\chardef\wfile=\ffile%
\immediate\write\ffile{\noexpand\medskip\noexpand\item{Fig.\ \the\figno. }
\reflabeL{#1\hskip.55in}\pctsign}\global\advance\figno by1\findarg}
\def\xfig{\expandafter\xf@g}\def\xf@g fig.\penalty\@M\ {}
\def\figs#1{figs.~\f@gs #1{\hbox{}}}
\def\f@gs#1{\edef\next{#1}\ifx\next\em@rk\def\next{}\else
\ifx\next#1\xfig #1\else#1\fi\let\next=\f@gs\fi\next}
\newwrite\lfile
{\escapechar-1\xdef\pctsign{\string\%}\xdef\leftbracket{\string\{}
\xdef\rightbracket{\string\}}\xdef\numbersign{\string\#}}

\def\writestop{\def\writestoppt{\immediate\write\lfile{\string\pageno%
\the\pageno\string\startrefs\leftbracket\the\refno\rightbracket%
\string\def\string\secsym\leftbracket\secsym\rightbracket%
\string\secno\the\secno\string\meqno\the\meqno}\immediate\closeout\lfile}}
\def\writestoppt{}\def\writedef#1{}
\def\seclab#1{\xdef #1{\the\secno}\writedef{#1\leftbracket#1}\wrlabeL{#1=#1}}
\def\subseclab#1{\xdef #1{\secsym\the\subsecno}%
\writedef{#1\leftbracket#1}\wrlabeL{#1=#1}}
\newwrite\tfile \def\writetoca#1{}
\def\leaderfill{\leaders\hbox to 1em{\hss.\hss}\hfill}
%   use this to write file with table of contents
\def\writetoc{\immediate\openout\tfile=toc.tmp
   \def\writetoca##1{{\edef\next{\write\tfile{\noindent ##1
   \string\leaderfill {\noexpand\number\pageno} \par}}\next}}}
%       and this lists table of contents on second pass

%
\catcode`\@=12 % at signs are no longer letters
%
%   Unpleasantness in calling in abstract and title fonts
\edef\tfontsize{\ifx\answ\bigans scaled\magstep3\else scaled\magstep4\fi}
\font\titlerm=cmr10 \tfontsize \font\titlerms=cmr7 \tfontsize
\font\titlermss=cmr5 \tfontsize \font\titlei=cmmi10 \tfontsize
\font\titleis=cmmi7 \tfontsize \font\titleiss=cmmi5 \tfontsize
\font\titlesy=cmsy10 \tfontsize \font\titlesys=cmsy7 \tfontsize
\font\titlesyss=cmsy5 \tfontsize \font\titleit=cmti10 \tfontsize
\skewchar\titlei='177 \skewchar\titleis='177 \skewchar\titleiss='177
\skewchar\titlesy='60 \skewchar\titlesys='60 \skewchar\titlesyss='60
\def\titlefont{\def\rm{\fam0\titlerm}% switch to title font
\textfont0=\titlerm \scriptfont0=\titlerms \scriptscriptfont0=\titlermss
\textfont1=\titlei \scriptfont1=\titleis \scriptscriptfont1=\titleiss
\textfont2=\titlesy \scriptfont2=\titlesys \scriptscriptfont2=\titlesyss
\textfont\itfam=\titleit \def\it{\fam\itfam\titleit}\rm}
 \ifx\answ\bigans\else scaled\magstep1\fi
\ifx\answ\bigans\def\abstractfont{\tenpoint}\else
\font\abssl=cmsl10 scaled \magstep1
\font\absrm=cmr10 scaled\magstep1 \font\absrms=cmr7 scaled\magstep1
\font\absrmss=cmr5 scaled\magstep1 \font\absi=cmmi10 scaled\magstep1
\font\absis=cmmi7 scaled\magstep1 \font\absiss=cmmi5 scaled\magstep1
\font\abssy=cmsy10 scaled\magstep1 \font\abssys=cmsy7 scaled\magstep1
\font\abssyss=cmsy5 scaled\magstep1 \font\absbf=cmbx10 scaled\magstep1
\skewchar\absi='177 \skewchar\absis='177 \skewchar\absiss='177
\skewchar\abssy='60 \skewchar\abssys='60 \skewchar\abssyss='60
\def\abstractfont{\def\rm{\fam0\absrm}% switch to abstract font
\textfont0=\absrm \scriptfont0=\absrms \scriptscriptfont0=\absrmss
\textfont1=\absi \scriptfont1=\absis \scriptscriptfont1=\absiss
\textfont2=\abssy \scriptfont2=\abssys \scriptscriptfont2=\abssyss
\textfont\itfam=\bigit \def\it{\fam\itfam\bigit}\def\footnotefont{\tenpoint}%
\textfont\slfam=\abssl \def\sl{\fam\slfam\abssl}%
\textfont\bffam=\absbf \def\bf{\fam\bffam\absbf}\rm}\fi
\def\tenpoint{\def\rm{\fam0\tenrm}% switch back to 10-point type
\textfont0=\tenrm \scriptfont0=\sevenrm \scriptscriptfont0=\fiverm
\textfont1=\teni  \scriptfont1=\seveni  \scriptscriptfont1=\fivei
\textfont2=\tensy \scriptfont2=\sevensy \scriptscriptfont2=\fivesy
\textfont\itfam=\tenit \def\it{\fam\itfam\tenit}\def\footnotefont{\ninepoint}%
\textfont\bffam=\tenbf \def\bf{\fam\bffam\tenbf}\def\sl{\fam\slfam\tensl}\rm}
\font\ninerm=cmr9 \font\sixrm=cmr6 \font\ninei=cmmi9 \font\sixi=cmmi6
\font\ninesy=cmsy9 \font\sixsy=cmsy6 \font\ninebf=cmbx9
\font\nineit=cmti9 \font\ninesl=cmsl9 \skewchar\ninei='177
\skewchar\sixi='177 \skewchar\ninesy='60 \skewchar\sixsy='60
\def\ninepoint{\def\rm{\fam0\ninerm}% switch to footnote font
\textfont0=\ninerm \scriptfont0=\sixrm \scriptscriptfont0=\fiverm
\textfont1=\ninei \scriptfont1=\sixi \scriptscriptfont1=\fivei
\textfont2=\ninesy \scriptfont2=\sixsy \scriptscriptfont2=\fivesy
\textfont\itfam=\ninei \def\it{\fam\itfam\nineit}\def\sl{\fam\slfam\ninesl}%
\textfont\bffam=\ninebf \def\bf{\fam\bffam\ninebf}\rm}
%
%---------------------------------------------------------------------
%

\hyphenation{anom-aly anom-alies coun-ter-term coun-ter-terms}
\def\inv{^{\raise.15ex\hbox{${\scriptscriptstyle -}$}\kern-.05em 1}}

\def\Dsl{\,\raise.15ex\hbox{/}\mkern-13.5mu D} %this one can be subscripted
\def\dsl{\raise.15ex\hbox{/}\kern-.57em\partial}

\def\tr{{\rm tr}} 
\font\bigit=cmti10 scaled \magstep1
 %pound sterling
\def\lspace{\ifx\answ\bigans{}\else\qquad\fi}
\def\lbspace{\ifx\answ\bigans{}\else\hskip-.2in\fi} % $$\lbspace...$$
\def\boxeqn#1{\vcenter{\vbox{\hrule\hbox{\vrule\kern3pt\vbox{\kern3pt
    \hbox{${\displaystyle #1}$}\kern3pt}\kern3pt\vrule}\hrule}}}
\def\mbox#1#2{\vcenter{\hrule \hbox{\vrule height#2in
        \kern#1in \vrule} \hrule}}  %e.g. \mbox{.1}{.1}
%   matters of taste
%\def\tilde{\widetilde} \def\bar{\overline} \def\hat{\widehat}
%
% some sample definitions
  %     curly letters

\def\vev#1{\langle #1 \rangle}

\def\darr#1{\raise1.5ex\hbox{$\leftrightarrow$}\mkern-16.5mu #1}
 %pound sterling

 %puts a small half in a displayed eqn
\def\roughly#1{\raise.3ex\hbox{$#1$\kern-.75em\lower1ex\hbox{$\sim$}}}

%\draftmode
\let\includefigures=\iftrue
\let\useblackboard=\iftrue
\newfam\black

%Figure Stuff
\includefigures
\message{If you do not have epsf.tex (to include figures),}
\message{change the option at the top of the tex file.}
\input epsf
\def\figin{\epsfcheck\figin}\def\figins{\epsfcheck\figins}
\def\epsfcheck{\ifx\epsfbox\UnDeFiNeD
\message{(NO epsf.tex, FIGURES WILL BE IGNORED)}
\gdef\figin##1{\vskip2in}\gdef\figins##1{\hskip.5in}% blank space instead
\else\message{(FIGURES WILL BE INCLUDED)}%
\gdef\figin##1{##1}\gdef\figins##1{##1}\fi}
\def\DefWarn#1{}
\def\figinsert{\goodbreak\midinsert}
\def\ifig#1#2#3{\DefWarn#1\xdef#1{fig.~\the\figno}
\writedef{#1\leftbracket fig.\noexpand~\the\figno}%
\figinsert\figin{\centerline{#3}}\medskip\centerline{\vbox{
\baselineskip12pt\advance\hsize by -1truein
\noindent\footnotefont{\bf Fig.~\the\figno:} #2}}
%\bigskip
\endinsert\global\advance\figno by1}
%%%
\else
\def\ifig#1#2#3{\xdef#1{fig.~\the\figno}
\writedef{#1\leftbracket fig.\noexpand~\the\figno}%
%\figinsert\figin{\centerline{#3}}\medskip
%\centerline{\vbox{\baselineskip12pt
%\advance\hsize by -1truein\noindent
%\footnotefont{\bf Fig.~\the\figno:} #2}}
%\bigskip\endinsert
\global\advance\figno by1} \fi

\input miniltx
\expandafter\def\expandafter\+\expandafter{\+}
\input url.sty
\input amssym

\def\id{{1 \kern-.28em {\rm l}}}

\def\K3{{\bf K3}}
\def\journal#1&#2(#3){\unskip, \sl #1\ \bf #2 \rm(19#3) }
\def\andjournal#1&#2(#3){\sl #1~\bf #2 \rm (19#3) }

\def\bar{\overline}
\def\hat{\widehat}
\def\ie{{\it i.e.}}

\def\tilde{\widetilde}

\def\frac#1#2{{#1\over#2}}

\def\hf{{\textstyle\half}}
\def\ket#1{|#1\rangle}

\def\vev#1{\langle#1\rangle}

\def\inbar{\,\vrule height1.5ex width.4pt depth0pt}
\def\IC{\relax\hbox{$\inbar\kern-.3em{\rm C}$}}
\def\IR{\relax{\rm I\kern-.18em R}}
\def\IP{\relax{\rm I\kern-.18em P}}

%
%%%%%%%%%%%%%%%%%%%%%%%%%%%%%%%%%%%%
%

%
\catcode`\@=11
\def\slash#1{\mathord{\mathpalette\c@ncel{#1}}}
\overfullrule=0pt

\def\EE{{\cal E}}
\def\FF{{\cal F}}

\def\HH{{\cal H}}
\def\II{{\cal I}}

\def\OO{{\cal O}}

\def\SS{{\cal S}}
\def\TT{{\cal T}}

\def\underrel#1\over#2{\mathrel{\mathop{\kern\z@#1}\limits_{#2}}}

\catcode`\@=12

%%%%%%%%%%%%%%%%%%%%%%%%%%%%%%%%%%%%%%%%%%%%%%%%%%%%%%%%%%%%%%

%

\def\ket#1{\left| #1\right\rangle}
\def\vev#1{\left\langle #1 \right\rangle}

\def\tr{{\rm tr}}

\def\exp{{\rm exp}}

%%%%%%%%%%%%%%%%%%%%%%%%%%%%%%%%%%%%%%%%%%%%%%%%%%%%%%%%%%%%%%
% new defs:

\def\p{{\partial}}

\def\ra{{\rightarrow}}

\def\hf{{\hat{f}}}
\def\hg{{\hat{g}}}

%ENTANGLEMENT CFT

%\CalabreseEW
\lref\CalabreseEW{
  P.~Calabrese, J.~Cardy and E.~Tonni,
  ``Entanglement negativity in quantum field theory,''
Phys.\ Rev.\ Lett.\  {\bf 109}, 130502 (2012).
[arXiv:1206.3092 [cond-mat.stat-mech]].
%%CITATION = arXiv:1206.3092%%
}

%\CalabreseNK
\lref\CalabreseNK{
  P.~Calabrese, J.~Cardy and E.~Tonni,
  ``Entanglement negativity in extended systems: A field theoretical approach,''
J.\ Stat.\ Mech.\  {\bf 1302}, P02008 (2013).
[arXiv:1210.5359 [cond-mat.stat-mech]].
%%CITATION = arXiv:1210.5359%%
}

%\VidalZZ
\lref\VidalZZ{
  G.~Vidal and R.~F.~Werner,
  ``Computable measure of entanglement,''
Phys.\ Rev.\ A {\bf 65}, 032314 (2002)..
}

%\CalabreseEZ
\lref\CalabreseEZ{
  P.~Calabrese, J.~Cardy and E.~Tonni,
  ``Entanglement entropy of two disjoint intervals in conformal field theory,''
J.\ Stat.\ Mech.\  {\bf 0911}, P11001 (2009).
[arXiv:0905.2069 [hep-th]].
%%CITATION = arXiv:0905.2069%%
}

%\CalabreseHE
\lref\CalabreseHE{
  P.~Calabrese, J.~Cardy and E.~Tonni,
  ``Entanglement entropy of two disjoint intervals in conformal field theory II,''
J.\ Stat.\ Mech.\  {\bf 1101}, P01021 (2011).
[arXiv:1011.5482 [hep-th]].
%%CITATION = arXiv:1011.5482%%
}

%\CardyMB
\lref\CardyMB{
  J.~L.~Cardy, O.~A.~Castro-Alvaredo and B.~Doyon,
  ``Form factors of branch-point twist fields in quantum integrable models and entanglement entropy,''
J.\ Statist.\ Phys.\  {\bf 130}, 129 (2008).
[arXiv:0706.3384 [hep-th]].
%%CITATION = arXiv:0706.3384%%
}

%\CalabreseEU
\lref\CalabreseEU{
  P.~Calabrese and J.~L.~Cardy,
  ``Entanglement entropy and quantum field theory,''
J.\ Stat.\ Mech.\  {\bf 0406}, P06002 (2004).
[hep-th/0405152].
%%CITATION = hep-th/0405152%%
}

%\PlenioZZ
\lref\PlenioZZ{
  M.~B.~Plenio and S.~Virmani,
  ``An Introduction to entanglement measures,''
Quant.\ Inf.\ Comput.\  {\bf 7}, 1 (2007).
[quant-ph/0504163].
%%CITATION = quant-ph/0504163%%
}

%REVIEWS

%\SolodukhinGN
\lref\SolodukhinGN{
  S.~N.~Solodukhin,
  ``Entanglement entropy of black holes,''
Living Rev.\ Rel.\  {\bf 14}, 8 (2011).
[arXiv:1104.3712 [hep-th]].
%%CITATION = arXiv:1104.3712%%
}

%\CalabreseQY
\lref\CalabreseQY{
  P.~Calabrese and J.~Cardy,
  ``Entanglement entropy and conformal field theory,''
J.\ Phys.\ A {\bf 42}, 504005 (2009).
[arXiv:0905.4013 [cond-mat.stat-mech]].
%%CITATION = arXiv:0905.4013%%
}

%\NishiokaUN
\lref\NishiokaUN{
  T.~Nishioka, S.~Ryu and T.~Takayanagi,
  ``Holographic Entanglement Entropy: An Overview,''
J.\ Phys.\ A {\bf 42}, 504008 (2009).
[arXiv:0905.0932 [hep-th]].
%%CITATION = arXiv:0905.0932%%
}

%ENTANGLEMENT HOLOGRAPHY

%\RyuBV
\lref\RyuBV{
  S.~Ryu and T.~Takayanagi,
  ``Holographic derivation of entanglement entropy from AdS/CFT,''
Phys.\ Rev.\ Lett.\  {\bf 96}, 181602 (2006). [hep-th/0603001].
%%CITATION = hep-th/0603001%%
}

%\RyuEF
\lref\RyuEF{
  S.~Ryu and T.~Takayanagi,
  ``Aspects of Holographic Entanglement Entropy,''
JHEP {\bf 0608}, 045 (2006).
[hep-th/0605073].
%%CITATION = hep-th/0605073%%
}

%\MyersXS
\lref\MyersXS{
  R.~C.~Myers and A.~Sinha,
  ``Seeing a c-theorem with holography,''
Phys.\ Rev.\ D {\bf 82}, 046006 (2010).
[arXiv:1006.1263 [hep-th]].
%%CITATION = arXiv:1006.1263%%
}

%\CasiniKV
\lref\CasiniKV{
  H.~Casini, M.~Huerta and R.~C.~Myers,
  %``Towards a derivation of holographic entanglement entropy,''
JHEP {\bf 1105}, 036 (2011).
[arXiv:1102.0440 [hep-th]].
%%CITATION = arXiv:1102.0440%%
}

%\MyersTJ
\lref\MyersTJ{
  R.~C.~Myers and A.~Sinha,
  ``Holographic c-theorems in arbitrary dimensions,''
JHEP {\bf 1101}, 125 (2011).
[arXiv:1011.5819 [hep-th]].
%%CITATION = arXiv:1011.5819%%
}

%\HartmanMIA
\lref\HartmanMIA{
  T.~Hartman,
  ``Entanglement Entropy at Large Central Charge,''
[arXiv:1303.6955 [hep-th]].
%%CITATION = arXiv:1303.6955%%
}

%\BelavinVU
\lref\BelavinVU{
  A.~A.~Belavin, A.~M.~Polyakov and A.~B.~Zamolodchikov,
  ``Infinite Conformal Symmetry in Two-Dimensional Quantum Field Theory,''
Nucl.\ Phys.\ B {\bf 241}, 333 (1984)..
%%CITATION = CERN-TH-3827%%
}

%\CasiniBW
\lref\CasiniBW{
  H.~Casini and M.~Huerta,
  ``A Finite entanglement entropy and the c-theorem,''
Phys.\ Lett.\ B {\bf 600}, 142 (2004).
[hep-th/0405111].
%%CITATION = hep-th/0405111%%
}

%\CasiniEI
\lref\CasiniEI{
  H.~Casini and M.~Huerta,
  ``On the RG running of the entanglement entropy of a circle,''
Phys.\ Rev.\ D {\bf 85}, 125016 (2012).
[arXiv:1202.5650 [hep-th]].
%%CITATION = arXiv:1202.5650%%
}

%\LiuEEA
\lref\LiuEEA{
  H.~Liu and M.~Mezei,
  ``A Refinement of entanglement entropy and the number of degrees of freedom,''
JHEP {\bf 1304}, 162 (2013).
[arXiv:1202.2070 [hep-th]].
%%CITATION = MIT-CTP-4336%%
}

%\LewkowyczNQA
\lref\LewkowyczNQA{
  A.~Lewkowycz and J.~Maldacena,
  ``Generalized gravitational entropy,''
JHEP {\bf 1308}, 090 (2013).
[arXiv:1304.4926 [hep-th]].
%%CITATION = arXiv:1304.4926%%
}

%\FaulknerYIA
\lref\FaulknerYIA{
  T.~Faulkner,
  ``The Entanglement Renyi Entropies of Disjoint Intervals in AdS/CFT,''
[arXiv:1303.7221 [hep-th]].
%%CITATION = arXiv:1303.7221%%
}

%\BarrellaWJA
\lref\BarrellaWJA{
  T.~Barrella, X.~Dong, S.~A.~Hartnoll and V.~L.~Martin,
  ``Holographic entanglement beyond classical gravity,''
JHEP {\bf 1309}, 109 (2013).
[arXiv:1306.4682 [hep-th]].
%%CITATION = SU-ITP-13-10%%
}

%\LitvinovSXA
\lref\LitvinovSXA{
  A.~Litvinov, S.~Lukyanov, N.~Nekrasov and A.~Zamolodchikov,
  ``Classical Conformal Blocks and Painleve VI,''
[arXiv:1309.4700 [hep-th]].
%%CITATION = RUNHETC-2013-15%%
}

%\MenottiKRA
\lref\MenottiKRA{
  P.~Menotti,
  ``On the monodromy problem for the four-punctured sphere,''
[arXiv:1401.2409 [hep-th]].
%%CITATION = IFUP-TH-2014-1%%
}

\lref\Zamolodchikov{
A.~B.~Zamolodchikov,
``Conformal symmetry in two-dimensional space: Recursion representation of conformal block,''
Theor. and Math. Phys. {\bf 73}, Issue 1, 1088 (1987)
}

%\HartmanOAA
\lref\HartmanOAA{
  T.~Hartman, C.~A.~Keller and B.~Stoica,
  ``Universal Spectrum of 2d Conformal Field Theory in the Large c Limit,''
[arXiv:1405.5137 [hep-th]].
%%CITATION = CALT-68-2889%%
}

%\HarlowNY
\lref\HarlowNY{
  D.~Harlow, J.~Maltz and E.~Witten,
  ``Analytic Continuation of Liouville Theory,''
JHEP {\bf 1112}, 071 (2011).
[arXiv:1108.4417 [hep-th]].
%%CITATION = SU-ITP-11-42%%
}

%\RangamaniYWA
\lref\RangamaniYWA{
  M.~Rangamani and M.~Rota,
  ``Comments on Entanglement Negativity in Holographic Field Theories,''
[arXiv:1406.6989 [hep-th]].
%%CITATION = arXiv:1406.6989%%
}

%\HeadrickZT
\lref\HeadrickZT{
  M.~Headrick,
  ``Entanglement Renyi entropies in holographic theories,''
Phys.\ Rev.\ D {\bf 82}, 126010 (2010).
[arXiv:1006.0047 [hep-th]].
%%CITATION = arXiv:1006.0047%%
}

%\HeadrickZDA
\lref\HeadrickZDA{
  M.~Headrick,
  ``General properties of holographic entanglement entropy,''
JHEP {\bf 1403}, 085 (2014).
[arXiv:1312.6717 [hep-th]].
%%CITATION = BRX-TH673%%
}

%\ChenKPA
\lref\ChenKPA{
  B.~Chen and J.~-J.~Zhang,
  ``On short interval expansion of Rényi entropy,''
JHEP {\bf 1311}, 164 (2013).
[arXiv:1309.5453 [hep-th]].
%%CITATION = arXiv:1309.5453%%
}

%\ChenDXA
\lref\ChenDXA{
  B.~Chen, J.~Long and J.~-j.~Zhang,
  ``Holographic Rényi entropy for CFT with W symmetry,''
JHEP {\bf 1404}, 041 (2014).
[arXiv:1312.5510 [hep-th]].
%%CITATION = arXiv:1312.5510%%
}

%\PerlmutterPAA
\lref\PerlmutterPAA{
  E.~Perlmutter,
  ``Comments on Renyi entropy in AdS$_3$/CFT$_2$,''
JHEP {\bf 1405}, 052 (2014).
[arXiv:1312.5740 [hep-th]].
%%CITATION = arXiv:1312.5740%%
}

%\BeccariaLQA
\lref\BeccariaLQA{
  M.~Beccaria and G.~Macorini,
  ``On the next-to-leading holographic entanglement entropy in $AdS_{3}/CFT_{2}$,''
JHEP {\bf 1404}, 045 (2014).
[arXiv:1402.0659 [hep-th]].
%%CITATION = arXiv:1402.0659%%
}

%\ChenKJA
\lref\ChenKJA{
  B.~Chen, F.~-y.~Song and J.~-j.~Zhang,
  ``Holographic Renyi entropy in AdS$_3$/LCFT$_2$ correspondence,''
JHEP {\bf 1403}, 137 (2014).
[arXiv:1401.0261 [hep-th]].
%%CITATION = arXiv:1401.0261%%
}

%ENTANGLEMENT NEGATIVITY

%\CalabreseMI
\lref\CalabreseMI{
  P.~Calabrese, L.~Tagliacozzo and E.~Tonni,
  ``Entanglement negativity in the critical Ising chain,''
J.\ Stat.\ Mech.\  {\bf 1305}, P05002 (2013).
[arXiv:1302.1113 [cond-mat.stat-mech]].
%%CITATION = arXiv:1302.1113%%
}

%\AlbaMG
\lref\AlbaMG{
  V.~Alba,
  ``Entanglement negativity and conformal field theory: a Monte Carlo study,''
J.\ Stat.\ Mech.\  {\bf 1305}, P05013 (2013).
[arXiv:1302.1110 [cond-mat.stat-mech]].
%%CITATION = arXiv:1302.1110%%
}

\lref\LeeVidal{
Y.~A.~Lee and G.~Vidal,
``Entanglement negativity and topological order,''
Phys.\ Rev. \ A  {\bf 88}, 042318 (2013)
[arXiv:1306.5711 [quant-ph]]
}

\lref\Castelnovo{
C.~Castelnovo
``Negativity and topological order in the toric code,''
Phys.\ Rev.\ A {\bf 88}, 042319 (2013)
[arXiv:1306.4990 [cond-mat.str-el]]
}

\lref\Molinaetal{
H.~Winterich, J.~Molina-Vilaplana and S.~Bose
``Scaling of entanglement between separated blocks in spin chains at criticality,''
Phys.\ Rev.\ A {\bf 80}, 010304(R) (2009)
[arXiv:0811.1285 [quant-ph]]
}

%%%%%%%%%%%%%%%%%%%%%%%%%%%%
NSF-KITP-14-096
\Title{}
{\vbox{\centerline{Conformal Blocks and Negativity at Large Central Charge}
\bigskip
\centerline{}
}}
\bigskip

\centerline{\it  Manuela Kulaxizi$^1$, Andrei Parnachev$^2$ and Giuseppe Policastro$^{3}$}
\bigskip
\smallskip
\centerline{${}^{1}$Physique Th\'eorique et Math\'ematique} 
\centerline{Universit\'e Libre de Bruxelles and International Solvay Institutes}
\centerline{Campus Plaine C.P. 231, B-1050 Bruxelles, Belgium}
\smallskip
\centerline{${}^{2}$Institute Lorentz for Theoretical Physics, Leiden University} 
\centerline{P.O. Box 9506, Leiden 2300RA, The Netherlands}
\smallskip
\centerline{${}^{3}$ Laboratoire de Physique Th\'eorique, Ecole Normale Sup\'erieure,} 
\centerline{24 rue Lhomond, 75231, Paris Cedex 05, France}
\centerline{(UMR du CNRS 8549)}
\smallskip

\vglue .3cm

\bigskip

\let\includefigures=\iftrue
\bigskip
\noindent
We consider entanglement negativity for two disjoint intervals in
1+1 dimensional CFT in the limit of large central charge.
As the two intervals get close, the leading behavior of negativity is
given by the logarithm of the conformal block where a set of approximately null
descendants appears in the intermediate channel.
We compute this quantity numerically and compare with existing analytic
methods which provide perturbative expansion in powers of the cross-ratio.
\bigskip

\Date{July 2014}

%\draftmode

\newsec{Introduction and summary}

\noindent 
Entanglement entropy has recently appeared in a variety 
of contexts ranging from AdS/CFT \refs{\RyuBV-\RyuEF} to entropic derivations
of the c-theorems \refs{\CasiniBW\CasiniEI-\LiuEEA}. Some reviews of the subject include \refs{\CalabreseQY\NishiokaUN\SolodukhinGN-\HeadrickZDA}.
%each focusing on different aspects of entanglement entropy.
%Reviews on the subject include
%(for a review see \CalabreseQY\ about entanglement entropy and 1+1 dimensional conformal field theory.
% \SolodukhinGN\ on aspects of entanglement entropy in higher dimensional systems and in relation to black %hole physics and
%\url{http://strings2013.sogang.ac.kr//main/?skin=video_OV1.htm} 
%It may sound surprising given the variety of applications, but there are circumstances where entanglement entropy fails to be a good measure of entanglement.
%Such is the case of a mixed state --
%where EE mixes quantum and classical correlations. 
%it is known for example that when the system is in a high temperature state entanglement entropy yields the same result as thermal entropy. Similar difficulties arise when measuring the entanglement in a pure state between two non-complementary parts. This is because the union of these two parts is generically in a mixed a state.
%One may wonder whether a quantity constructed from entanglement entropy, referred to as the mutual %information, can be succesfully used in such cases. It turns out however that the mutual information does %not have all the necessary properties which constitute a proper entanglement measure \PlenioZZ.
 An alternative measure of entanglement, called entanglement negativity, has been introduced in  \VidalZZ\ and further discussed in \CalabreseEW\ and \CalabreseNK (see also \refs{\Molinaetal\AlbaMG\CalabreseMI \Castelnovo-\LeeVidal}). The entanglement negativity appears to be a good entanglement measure for systems in a mixed state.
It was additionally shown that in 1+1 dimensional conformal field theories (CFTs) the value
of the entanglement negativity for a single interval is proportional to the central charge
of the theory, similarly to entanglement entropy. 
Hence, entanglement negativity might be useful for defining a proper measure of
degrees of freedom, something that entanglement entropy is having 
some issues with, especially in four and higher space-time dimensions \LiuEEA.

The holographic prescription for computing entanglement entropy \refs{\RyuBV-\RyuEF} has given
us a simple and efficient computational technique.
A derivation based on the replica trick was provided in \LewkowyczNQA.
The key property was the $n\rightarrow1$ limit of the replica index: the contribution
proportional to $(n-1)$ is localized on minimal surfaces in AdS. 
Entanglement negativity also involves $n\rightarrow1$, but is supposed
to be finite in this limit, thanks to the non-trivial analytical continuation from even $n$
(to be reviewed below).
Hence, it is non clear at present whether a nice holographic formula for 
entanglement negativity exists.
(See \RangamaniYWA\ for a recent discussion.)

In two space-time dimensions conformal symmetry is much more powerful 
than in higher dimensions, and one can recover some of the holographic results
for entanglement entropy
by going to the limit of large central charge and making some mild assumption
about the spectrum of the operators and the OPE coefficients \HartmanMIA\
(see also \refs{\HeadrickZT\FaulknerYIA\BarrellaWJA -\HartmanOAA} and for related work  \refs{\ChenKPA\ChenDXA\PerlmutterPAA\BeccariaLQA-\ChenKJA}).
One may ask whether a similar approach can be useful in the computation of 
entanglement negativity, and if the holographic prescription can be
inferred from it.

In this paper we attempt to answer the first question by computing entanglement
negativity for two disjoint intervals in 1+1 dimensional CFTs in the limit of large
central charge $c$.
Technically we need to compute a four-point function of twist operators in an arbitrary CFT.
The difference from the entanglement entropy computation arises from the 
fact that in the limit of close intervals the two  operators that come together 
are not the twist and its inverse, but rather two identical twist operators.
As a result, the leading contribution in this limit does not come from the 
identity, but from a non-trivial twist operator.
As in \HartmanMIA, we assume that the computation of the conformal block,
associated with this operator is sufficient to give the full result for the
negativity (unlike \HartmanMIA, we do not have a non-trivial holographic
check of this assumption).

Generally there exists a number of ways to compute conformal blocks at large $c$
as an expansion in powers of the cross-ratio.
However the conformal block that appears in the result for negativity is particularly non-trivial:
the internal operator has a descendant which is null in the limit of infinite central charge.
This shows up as a divergence in the next to next to leading term in the expansion.
In situations like this one generally expects a series of divergent terms which can be
resumed. 
This is technically a hard problem, so instead we reformulate the problem in terms 
of a auxiliary differential equation with fixed monodromy,  and solve it numerically.
Interestingly, we find two solutions; we pick up the dominant one, while the other is exponentially
suppressed in the limit of large central charge.

The rest of the paper is organized as follows. 
In the next Section we review the definition of entanglement negativity and
how it can be computed in a 1+1 dimensional CFT using the replica trick.
In Section 3 we describe our result.
We discuss it in Section 4. In the Appendix we explore some analytic techniques for computing the conformal bock as an expansion in powers of the cross-ratio. 

%PERHAPS ADD:
%Holographic calculation ala Myers in three dimensions.
%Result from Myers about Renyi entropy which shows negativity is not proportional to a or c.
%Extend our result to higher dimensions for two strips?
%Consider the case of two spheres, one inside the other. 
%Different c2 and easier limit, perhaps an explicit map exists.

\newsec{Negativity in a 1+1 dimensional CFT: a review}

\noindent 
In this Section we briefly review the results of \CalabreseEW\ and \CalabreseNK.
We start by recalling the definition of negativity \VidalZZ, used as a measure of entanglement
between two quantum subsystems $A_1$ and $A_2$ whose Hilbert spaces are denoted by $\HH_1$ and $\HH_2$ below.
Denote by $|e_i^{(1)}\rangle$ and   $|e_j^{(2)}\rangle$ the bases of $\HH_1$ and $\HH_2$  respectively.
The partial transpose of the density matrix $\rho$ in $\HH_1 \cup \HH_2$ is defined as
\eqn\transpose{  \langle e_i^{(1)}\, e_j^{(2)} | \rho^{T_2}  | e_k^{(1)}\, e_l^{(2)} \rangle =   \langle e_i^{(1)}\, e_l^{(2)} | \rho | e_k^{(1)}\, e_j^{(2)} \rangle \,.  } 
Then the entanglement negativity is simply defined as 
\eqn\negdef{   \EE \equiv \ln \tr | \rho^{T_2}| \,, }
where $\tr | \rho^{T_2}|$ denotes the sum of the absolute values of the eigenvalues of $\rho^{T_2}$. The transposed matrix is normalized, $tr(\rho^{T_2})=1$, but it is not necessarily a density matrix, since it can have negative eigenvalues that are counted by the negativity, hence the name. 

As explained in \CalabreseEW, one can compute the value of $\EE$ using the replica trick.
%, a method which was very useful in computing entanglement entropy. 
To see how, it is useful to express the trace of $\rho^{T_2}$ in terms of its eigenvalues $\lambda_i$
\eqn\tracea{\tr|{\rho^{T_2}}|=\sum_i |\lambda_i|=\sum_{\lambda_i>0}|\lambda_i|+\sum_{\lambda_i<0}|\lambda_i|\,.} 
and subsequently consider the trace of integer powers of $\rho^{T_2}$. It is evident that the dependence of  $\tr{(\rho^{T_2})^n}$ on $|\lambda_i|$ will differ for odd and even integer values of $n$, \ie,
\eqn\tracen{\tr{(\rho^{T_2})^n}=\sum_i\lambda_i^n=\left\{\eqalign{&\sum_{\lambda_i>0}|\lambda_i|^n+\sum_{\lambda_i<0}|\lambda_i|^n ,\qquad n:even\cr
&\sum_{\lambda_i>0}|\lambda_i|^n-\sum_{\lambda_i<0}|\lambda_i|^n,\qquad n:odd
}
\right. \,.}
 The authors of \CalabreseEW\ observed that if one formally sets $n=1$ in the upper line of eq. \tracen,
one obtains exactly eq.\tracea. This is not true for the case of odd integer $n$, where the lower line of eq.\tracen\ simply yields the normalization condition $\tr\rho^{T_2}=1$. 
Disregarding issues with the existence of a unique analytic continuation, entanglement negativity can thus be obtained by analytically continuing the {\it even} sequance of integer numbers $n$ and then taking the limit $n\rightarrow 1$,
\eqn\eereplica{\EE=\lim_{n_e\rightarrow 1}\ln\tr{(\rho^{T_2})^{n_e}}\,,}
where, following the conventions of \CalabreseEW\ we defined $n_e\equiv 2m$ for integer $m$.

\noindent A consistency check for the replica trick method is performed in \CalabreseEW\ where $\EE$ is evaluated for the case of a bipartite system $\HH_A\times\HH_B$ in a pure state $\ket{\Psi}$ and the result is shown to agree with \VidalZZ, \ie,  entanglement negativity is equal to the Renyi entropy of order $1/2$ 
\eqn\eepure{\EE=2\ln{[\tr{(\rho_A^{1\over 2}})]}\,.}

The replica trick has been very useful in obtaining results for entanglement entropy in 2d CFTs, see for example \CalabreseQY\ and references therein. The success was due to expressing integer powers of the trace of the reduced density matrix in terms of correlation functions of twist operators 
%(enforcing the appropriate boundary conditions in the path integral formalism) 
which are fixed by conformal invariance up to a few independent parameters.
It turns out \CalabreseEW\ that one can similarly express $\tr[(\rho^T)^n]$ in terms of twist operators
and thus compute $\EE$ in a few simple cases in $1+1$ dimensional CFTs. However, the situation for entanglement negativity is slightly more complicated than that for entanglement entropy, due to the nature of the analytic continuation (from even $n$ to $n\rightarrow 1$ in \eereplica ).
%In a few simple cases in $1+1$ dimensional CFTs, it is possible to compute $\EE$ by combining the replica %trick with the methods of 2d CFT. 

The simplest non-trivial configuration to consider involves two disjoint intervals,
\eqn\defints{  A_1 = (z_1,z_2), \qquad A_2 = (z_3, z_4), \qquad z_2 < z_3  \quad .}    
%\eqn\defints{  A_1 = (u_1,\upsilon_1), \qquad A_2 = (u_2, \upsilon_2), \qquad \upsilon_1 < u_2  }  
The authors of \CalabreseEW\ showed that in this case we can identify integer powers of the transpose of the reduced density matrix with the following four-point function of twist operators
\eqn\fpa{\tr(\rho_{A_2}^T)^n=\vev{\TT_n(z_1)\bar{\TT}_n(z_2)\bar{\TT}_n(z_3)\TT_n(z_4)}\,}
where $\TT_n,\,\bar{\TT_n}$ have conformal dimensions (see for example \CalabreseEU ):
\eqn\twistcd{h_{\TT_n}=h_{\bar{\TT}_n}={c\over 24}\left(n-{1\over n}\right)\equiv h\,.}
Following eq. \eereplica\ entanglement negativity can thus be obtained from the analytic continuation of the logarithm of this four-point function for even $n$, 
\eqn\eefp{\EE=\lim_{n_e\rightarrow 1}\,\ln{\left[ \vev{\TT_{n_e}(z_1)\bar{\TT}_{n_e}(z_2)\bar{\TT}_{n_e}(z_3)\TT_{n_e}(z_4)}\right]}}

Conformal invariance implies that entanglement negativity is a function of the cross-ratio,
\eqn\defx{  x = { (z_2-z_1) (z_4 - z_3) \over (z_3-z_1) (z_4-z_2) } ={\ell_1\ell_2\over (\ell_1-(z_3-z_2))(\ell_2+(z_3-z_2))}  }
%\eqn\defx{  x = { (\upsilon_1-u_1) (\upsilon_2 - u_2) \over (u_2-u_1) (\upsilon_2-\upsilon_1) }   }
where we defined $\ell_1$ and $\ell_2$ to be the length of the intervals $A_1$ and $A_2$ respectively.
An interesting limit involves two intervals that are taken to be very close to each other.
In this adjacent interval limit, $x\ra 1$ and  
\eqn\eeadj{  \EE   \simeq -{c\over4} \log (1-x)}
Another limit involves taking the two intervals far away from each other, $x\ra0$.
As explained in \CalabreseNK, in this limit negativity is non-perturbatively small:
all coefficients in front of powers of $x$ vanish identically.

\newsec{Entanglement negativity in the limit of large central charge}

\noindent In this section, we will review the general characteristics of four point functions of primary operators in the limit of large central charge (for a detailed recent review see e.g. \HarlowNY). We will then proceed to study the four-point function through which entanglement negativity is defined (eq. \eefp ) and compute $\EE$ using some mild assumptions about the behavior of the ope coefficients and the spectrum of operators. The assumptions are exactly the ones used in \HartmanMIA\ and which reproduced the holographic result for the entanglement entropy of disjoint intervals.

%\eqn\twistcd{h_{\TT_n}=h_{\bar{\TT}_n}={c\over 24}\left(n-{1\over n}\right)\equiv h\,.}
%One is further supposed to take a limit $n\ra 1$
%\eqn\negrep{  \EE=\lim_{n_e\ra 1} \,  \vev{\TT_n(z_1)\bar{\TT}_n(z_2)\bar{\TT}_n(z_3)\TT_n(z_4)}_{even\,\, %n_e} }
%which is highly non-trivial 
%because \enfourp\  behaves differently for even and odd $n$.
%It is instructive to compare \enfourp\  with entanglement Renyi entropy of two disjoint intervals,
%which is given by another four point function, namely %$\vev{\TT_n(z_1)\bar{\TT}_n(z_2)\TT_n(z_3)\bar{\TT}_n(z_4)}$.
%Entanglement negativity depends solely on the cross-ratio $x$ defined in \defx. 

Consider the four point function of primary operators $\vev{\OO_1(z_1)\OO_2(z_2)\OO_3(z_3)\OO_4(z_4)}$.
Conformal invariance allows us to set $z_1=0,\,z_2=x,\,z_3=1,\,z_4=\infty$ and focus on  $\vev{\OO_1(0)\OO_2(x)\OO_3(1)\OO_4(\infty)}$. Moreover, it implies that any four point function of primary 
operators $\OO_i$ can be decomposed into conformal blocks
\eqn\fourpg{ \vev{\OO_1(0)\OO_2(x)\OO_3(1)\OO_4(\infty) }=
 \sum_p a_p\FF(c,h_p,h_i,x) \bar{\FF} (c,\bar{h}_p,\bar{h}_i,\bar{x})\,,}
where $(h_i,\bar{h}_i)$ denote the conformal dimensions of the operators $\OO_i$ and 
the summation is over primary operators $\OO_p$ with conformal dimension $(h_p,\bar{h}_p)$.
An analytic expression for $\FF(c,h_p,h_i,x)$ is known only for particular values of the parameters. In general, it is computed in the form of a series expansion in powers of $x$ or $(1-x)$ depending on whether \fourpg\ is a $s-$ or $t-$channel decomposition into conformal blocks. 

However, in the limit of large central charge $c\gg 1$ and fixed $\left({h_p\over c},\,{h_i\over c}\right)$, the conformal blocks acquire a simple exponential from \refs{\Zamolodchikov-\BelavinVU}
\eqn\cfexp{\FF(c,h_p,h_i,x)\sim \exp{\left[-{c\over 6} f\left({h_p\over c},{h_i\over c},x\right)\right]}\,.}
The function $f\left({h_p\over c},{h_i\over c},x\right)$ is then determined by the monodromy properties of a second order ordinary differential equation,
\eqn\difeq{\psi''(z)+T(z)\psi(z)=0 }
where
\eqn\defT{ T(z)=\sum_{i=1}^{i=4} \left({6 h_i\over c(z-z_i)^2}-{c_i\over z-z_i}\right)}
is related to the semiclassical, \ie, large $c$, stress energy tensor through the Ward Identity $\vev{T(z)\OO_1(z_1)\OO_2(z_2)\OO_3(z_3)\OO_4(z_4)}=\sum_{i=1}^4 \left({h_i\over (z-z_i)^2}+{\p_i\over z-z_i}\right) \,\vev{\OO_1(z_1)\OO_2(z_2)\OO_3(z_3)\OO_4(z_4)}$.
The constants $c_i$ in \difeq\ are called accessory parameters. Requiring $T(z)$ to behave like $z^{-4}$ as $z\rightarrow \infty$ leads to
\eqn\acceqs{\sum_i c_i=0,\qquad \sum_i c_i z_i-{6 h_i\over c}=0,\quad \sum_ic_i z_i^2-{12 h_i\over c}z_i=0\,.}
These equations determine three out of the four accessory coefficients so that, after setting $(z_1,z_2,z_3,z_4)=(0,x,1,\infty)$, $T(z)$ becomes
\eqn\Tz{T(z)={6 h_1\over cz^2}+{6h_2\over c(z-x)^2}+
{6 h_3\over c(z-1)^2}+{6(h_1+h_2+h_3-h_4)\over cz(1-z)} -{c_2 x(1-x)\over z(z-x)(1-z)}\,. }
To specify $c_2(x)$ one is then instructed to consider the trace of the monodromy matrix $M$
of the solutions $\psi_{1,2}(z)$ of \difeq\ around a path enclosing points $(0,x)$ for the $s-$channel or $(x,1)$ in the $t-$channel and require that 
\eqn\traceM{\tr M=-2 \cos{\pi\Lambda_p}, \qquad h_p={c\over 24}(1-\Lambda_p^2)}
Finally, the semiclassical conformal block $f(c,h_p,h_i,x)$ is related to $c_2(x)$ via
\eqn\fdifeq{{\p f\over\p x}=c_2(x)}
Assuming now that $f(c,h_p,h_i,x)$ is known for all operators $\OO_p$, one can compute the four point function \fourpg\ in the semiclassical limit via 
\eqn\fourpgf{\vev{\OO_1(0)\OO_2(x)\OO_3(1)\OO_4(\infty) }\simeq\sum_p a_p \exp{\left[-{c\over 6} f\left({h_p\over c},{h_i\over c},x\right)-{c\over 6}\bar{f}\left({h_p\over c},{h_i\over c},x\right)\right]}\,.} 

%Unfortunately, it is not known in practice how to analytically solve the monodromy problem of eq.\difeq\ 
%and thus compute the semiclassical conformal block $f({h_p\over c},{h_i\over c},x)$. There are however some cases where a great simplification can occur; namely, when the sum in \fourpgf\ is dominated by a specific operator $\OO_p$ for which 
%the monodromy problem is analytically tractable. Such is the case of the four point function $\vev{\TT_n(u_1)\bar{\TT}_n(v_1)\TT_n(u_2)\bar{\TT}_n(v_2)}$ which determines the entanglement Renyi entropy \HartmanMIA.
%It was argued in \HartmanMIA that in the semiclassical limit 
%(and as long as the conformal dimensions $h_i$ are all equal -- ??), 
%the leading contribution to the sum in \fourpgf\ comes from the "zeroth" order conformal block, \ie, the operator $\OO_p$ with ${h_p\over c}=0$. {\bf  MORE} .....
%In higher dimensional CFTs this is just the identity operator but in $d=2$ dimensions this is not true. In fact, the stress energy tensor operator has ${h_p\over c}=0$.
%From \traceM\ ${h_p\over c}=0$ translates to the requirement of the trivial monodromy for the solutions of \difeq.
%The trivial monodromy problem is fortunately rather simple and was solved analytically in \HartmanMIA.

It was argued in \HartmanMIA\ that, under some mild assumptions, in the semiclassical regime the dominant contribution to \fourpgf\ comes from the conformal block associated to operators $\OO_p$ of the lowest dimension $h_p\over c$. In other words, \fourpgf\ can be well approximated by the first term in the sum.
%A FEW WORDS ABOUT ASSUMPTIONS.

The analysis above was used in \HartmanMIA\  to compute the entanglement entropy of two disjoint intervals in the limit of large central charge (recall that the entanglement entropy of two disjoint intervals is related to $\vev{\TT_n(0)\bar{\TT}_n(x)\TT_n(1)\bar{\TT}_n(\infty)}$). Ref. \HartmanMIA\ showed that the light operators to consider are the dimension zero operators which correspond to trivial monodromy according to \traceM. The result obtained precisely matched the one derived earlier from holography \RyuEF.

%A way to understand the trivial monodromy requirement in the $s-$channel is to study the limit %$x\rightarrow 0$. Starting from the four point function and taking the limit $x\rightarrow 0$ results in 
%\eqn\fourptptwo{\lim_{x\rightarrow %0}\vev{\TT_n(0)\bar{\TT}_n(x)\TT_n(1)\bar{\TT}_n(\infty)}=\vev{\II(0)\TT_n(1)\bar{\TT}_n(\infty)}=
%\vev{\TT_n(1)\bar{\TT}_n(\infty)}} 
%where we used the fact that  ${\bar\TT}_n=\TT_n^\dagger=\TT_{-n}$ and thus $\TT_n\bar{\TT}_n(0)\equiv %\II$ is equal
%to the identity operator (see e.g. \CardyMB\ ). 
%Solving for the accessory parameters either from the three point function or the two point function in %\fourptptwo\ yields
%\eqn\TTzeror{T(z)={6 h\over c}{(z-1)^2}\, ,}
%with $h$  defined in \twistcd.
%The solutions of eq. \difeq\  with $T(z)$ given by \TTzeror\ behave like exponentials in the vicinity of 
%$z\simeq 0$ and thus have trivial monodromy around the point $z=0$.

%Although the above discussion was resticted to the s-channel similar analysis can be performed in 
%the t-channel $x\simeq 1$. Solutions of \difeq\ at $x=1$ for the entanglement Renyi entropy have trivial %monodromy 
%in accordance with \HartmanMIA. 

We would like to perform a similar analysis to compute the entanglement negativity for two disjoint intervals.  
As mentioned in the previous section, we should focus on the following four point function of twist operators
\eqn\enfourp{\vev{\TT_{n_e}(z_1)\bar{\TT}_{n_e}(z_2)\bar{\TT}_{n_e}(z_3)\TT_{n_e}(z_4)} }
with conformal dimensions given by eq.\twistcd.
We shall again assume that the conformal block with the smalllest conformal dimension provides the leading contribution to the sum \fourpgf. So the problem amounts to identifying the operator $\OO_p$ with the smallest dimension.

Let us consider the limiting cases $x=0$ and $x=1$.
For $x\rightarrow 0$ the four point function of \enfourp\ leads to
\eqn\fourptpa{\lim_{x\rightarrow 0}\vev{\TT_n(0)\bar{\TT}_n(x)\bar{\TT}_n(1)\TT_n(\infty)}=
\vev{\II(0)\bar{\TT}_n(1)\TT_n(\infty)}=\vev{\bar{\TT}_n(1)\TT_n(\infty)}\,,}
which implies trivial monodromy for the solutions of \difeq\ (note that this expression does 
not depend on whether $n$ is even or odd). The two-point function in \fourptpa\ is proportional to $(n-1)$ and thus vanishes in the limit $n\rightarrow 1$. Entanglement negativity is non-perturbatively small for $x=0$ as explained in \CalabreseNK.

For $x= 1$ on the other hand,
\eqn\fourptpb{\lim_{x\rightarrow 1}\vev{\TT_n(0)\bar{\TT}_n(x)\bar{\TT}_n(1)\TT_n(\infty)}=
\vev{\TT_n(0)\bar{\TT}_n^2(1)\TT_n(\infty)} \,,}
The conformal dimension of $\bar{\TT}_n^2$ was found in \refs{\CalabreseEW-\CalabreseNK} to be
\eqn\tsquaredim{\hat{h}_n\equiv h_{\TT^2_n}=\left\{ \eqalign{ &{c\over 24}\,\,\left(n-{1\over n}\right),\qquad n:odd\cr
&{c\over 12}\,\,\left({n\over 2}-{2\over n}\right),\qquad n:even} \right. }
To compute the negativity we are instructed to consider \fourptpb\ for $n$ even, analytically continue and take the limit $n\rightarrow 1$. 
In this limit the conformal dimension of the $\TT^2_n$ operator is
\eqn\limdelta{  {\hat h} \equiv\lim_{n_{even} \ra 1} {\hat h}_n = -{c\over 8}   }
For the three point function, one can explicitly determine all accessory parameters in \acceqs.
Solving \acceqs\ for the three point function in \fourptpb\ with even $n$ we obtain
\eqn\TTthreen{T(z)={6 h\over c}\, {z^2+(a-2)z+1\over z^2(z-1)^2},\qquad a\equiv { {\hat{h}}\over h}=2{{n\over 2}-{2\over n}\over n-{1\over n} }\,, }
and $h$ is defined in \twistcd. The differential equation \difeq\ can be solved analytically in the neighborhood of $z\sim 1$. The solutions read
\eqn\solthree{\psi_{\pm}(z)\simeq (z-1)^{{1\over 2} \left(1\pm \sqrt{1-24 {\hat{h}\over c}}\right)}\, ,}
and the trace of the monodromy matrix is $\tr{M}_{(x=1)}=-2 \cos{\left[\pi \sqrt{1+{4\over n}-n}\right]}$.
When $n\rightarrow 1$ the  trace of the monodromy matrix reduces to 
\eqn\trmxone{\tr{M}_{(x,1)}|_{n\rightarrow 1}=-2}
We could also deduce this monodromy by setting  $n=1$ in \tsquaredim\ evaluated for even $n$.
This corresponds to $\Lambda_p^2=4$ according to \traceM\ which gives rise to \trmxone.
%
%\ifig\loc{The log-log plot of $x c_2(x)$ computed numerically. The power is very close to $x^{3\over4}$.}
%{\epsfxsize3.3in\epsfbox{loglog.eps}}
%

In summary, to find the negativity in the vicinity of $x=1$ we need to solve eq. \difeq\
with 
\eqn\Tzneg{  T(z) = -{c_2 x (1-x) \over z (z-x) (1-z)  }  }
and impose the monodromy condition \trmxone\ which corresponds to an intermediate operator of conformal dimension $h_p={\hat h}$ given in \limdelta.
We also need to ensure 
\eqn\limctwo{   c_2 \simeq -{3\over 4} \, {1\over 1-x}, \quad x\approx 1}
to recover \TTthreen\ when $x=1$.
Entanglement negativity is then obtained from \fourpg, 
%\cfexp, \fdifeq:
\eqn\negint{  {\p \EE\over \p x} = {c\over 3}  c_2(x)  }
It is relatively easy to recover 
 \eeadj\  when $x\ra 1$. 
 In this limit the equation becomes
 \eqn\eqTlim{   \psi''(z) + {c_2 (1-x) \over (z-1)^2  } \psi(z) = 0 }
and the choice of the accessory parameter \limctwo\ gives rise to the solutions $\psi_1(z) \simeq (z-1)^{-{1\over2}}, \, \psi_2(z) \simeq (z-1)^{3\over2}$ with
the monodromy \trmxone. 
Unfortunately we could not solve the general problem analytically,
but we could do numerical integration.
Below we describe the operational procedure.
We start by rewriting the differential equation in the form
\eqn\difeqb{    h'(z) +h^2(z) +T(z) = 0}
where $\psi(z) = \exp(\int_{z_0}^z dz' h(z'))$.
To compute the monodromy, we integrate over a circle $z=1+r_0 e^{2\pi i t},\, t\in(0,1)$.
We make sure the contour encircles both $1$ and $x$, $r_0>(1-x)$.
Eq. \difeqb\ then reads
\eqn\difeqc{ {\dot h\over 2\pi i r_0 e^{2\pi i t} } +h^2 +T = 0}
where the dot denotes differentiation wrt $t$.
We consider two independent solutions specified by the value of $h(t=0)$:
\eqn\twosols{   h_1(t=0) =  0, \qquad h_2(t=0)=1}
This is translated to the following initial conditions at $z_0=1+r_0$:
\eqn\initone{   \psi_1(z_0) = 1,\qquad \psi_1'(z_0) = h_1(t=0) =0   }
and
\eqn\inittwi{  \psi_2(z_0) =1, \qquad \psi_2'(z_0) = h_2(0) =1}
On the other hand, we can integrate eq. \difeqc\ numerically to obtain
the values of $\psi_{1,2}$ at $t=1$.
Simple algebra then leads to the following expression for the monodromy trace:
\eqn\montrh{  \tr M_{(x,1)} = (1-a_1^{(1)}) e^{a_2^{(1)}} +a_1^{(2)} e^{a_2^{(2)}}  }
where
\eqn\defa{   a_1^{(i)} = h_i(t=1), \qquad a_2^{(i)} = 2\pi i r_0 \int_0^1 dt e^{2\pi i t} h_i(t) , \qquad i=1,2  }
We perform numerical integration. 
Independence of $r_0$ and limiting behavior \limctwo\ serve as simple checks.
Another simple check involves comparing the solution of the monodromy problem with 
the expansion of the conformal block in terms of the cross-ratio \BelavinVU:
\eqn\confblock{ F(h_p,y) = y^{h_p} \left(1+{1\over2}h_p y + {h_p (h_p+1)^2\over 4 (2 h_p+1)} y^2
         +{h_p^2 (1-h_p)^2\over 2 (2h_p+1)  ( c (2h_p+1) + 2 h_p (8h_p-5))  } y^2     +\ldots \right) }
Here $y= 1-x$ and $h_p$ is the dimension of the intermediate operator, what is called $\Delta'$ in \BelavinVU.
We also set $\Delta$ of \BelavinVU\ to zero.
It is now clear that for a generic $h_p\simeq a c$ the first three terms in the bracket
correspond to exponentiation
\eqn\fexpbpz{  \exp\left( c( a \log y + {a\over2}  y+ b y^2\ldots ) \right)   \simeq    
        y^\Delta \left( 1+   {c a\over2}  y  +  {c^2 a^2\over 8} y^2 +   c b y^2 +\ldots\right) }
Comparing this with \confblock\ implies
\eqn\bres{  b =   a \left( {3\over 16} +  {a\over 8 (1+ 8 a)}   \right) }
which also agrees with  \refs{\LitvinovSXA-\MenottiKRA}.
We verified that our numerics reproduces the first three terms in the logarithm of the
conformal block \cfexp\ for arbitrary generic values of $a$.

\ifig\loc{${\tilde C} = 1- C_2(y)/C_2(0)$ as a function of $y$.
There are two solutions (dots) approximated by eq. (3.32). }
{\epsfxsize3.3in\epsfbox{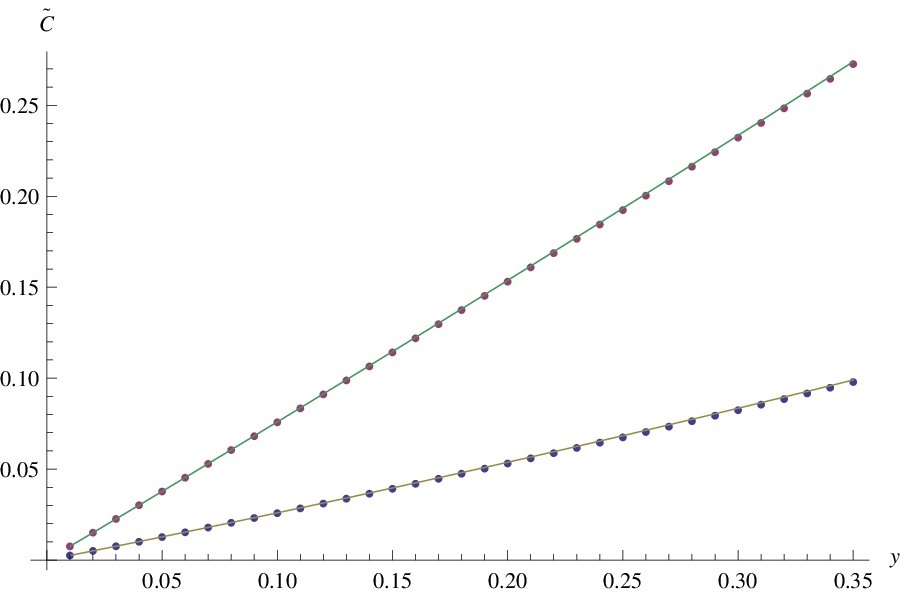}}       

\noindent Having checked our numerics we proceed to solve for the case of interest, $h_p=-{c\over 8}$. 
Contrary to the generic cases discussed above, we find {\it two} solutions (Fig.1). They are approximated by
\eqn\solapp{  C_2^\pm(y) = y (1-y) c_2( y) =- {3\over4}  \left(  1 -\left( {1\over 2} \pm{1\over 4}\right) y  +\ldots   \right)   }
At large central charge we are instructed to choose the dominant solution, 
which corresponds to $C_2^-$.
We then recover negativity by integrating \negint.
For sufficiently small $x$ there is a phase transition to the other branch,
where negativity is simply zero.
(For small $x$ the computation of the four-point function is identical to the one performed for entanglement entropy,
and there is a factor of $(n-1)$ which vanishes in the $n\ra1$ limit.)
We have checked that the two solutions never cross each other within the accuracy of our numerics.

\newsec{Discussion}
We computed a conformal block which at large central charge gives rise to 
the value of the negativity for two disjoint intervals. We performed a numerical computation because we could not determine the accessory parameter analytically by solving the differential equation \difeq. 
It would be interesting to compare our result to the analytic expansion \confblock.
However, precisely for $h_p=a c ={\hat h}= -c/8$ (which corresponds to $n=1$ in \tsquaredim, and which is 
the leading operator that appears in the OPE expansion of $\TT$ and $\bar\TT$) 
we encounter a complication.
In this case the last term in \confblock\ gets
enhanced by a factor of $c$ and becomes $\OO(c^2 y^2)$.
This is related to the fact that our operator has a level two descendant which
is null at leading order in central charge.
Hence, the large $c$ limit and the small $x$ limits do not have to commute in this
case.

One may wonder whether an analytic method for solving the Heun equation \difeq\ order by order in $y$ could be applied here (no issue with the order of limits in this case). A priori, there is no reason to doubt that this can be done. However, as we show in the appendix one finds the same difficulties as with the standard conformal block expansion. Namely, a divergent quadratic term. We believe that this hints to some non-analytic behavior of the conformal block.
Indeed as explained above, the complication is traced to the fact that the intermediate operator has a null
descendant at level two at leading order in the large $c$ expansion.
(This is true for all operators which correspond to $tr M = \pm 2$ with the exception
of the identity). In fact, there is an infinite tower of null states of even levels and 
and it is natural to expact similar divergences to appear at higher orders in the expansion
in powers of $y^2$; the simplest non-analytic behavior resulting from the resummation could be $\pm\sqrt{y^2} =\pm y$, which would explain the origin of the $\pm {y/4}$ term in \solapp.
(From eq. \fexpbpz\ the coefficient of the linear term is ${3\over 8}$, in agreement with  \LitvinovSXA\ and \MenottiKRA. )

A related question is the validity of the derivation of \difeq\ in our case.
This derivation has been recently reviewed in the Appendix D of \HarlowNY\ and we do not find 
any fundamental differences when a null descendant is present in the intermediate
channel at the leading order in the central charge expansion.
One may wonder whether exponentiation  of the conformal block \cfexp\ is still 
valid.
It would be nice to address this question.

The method used in this paper does not allow the computation of the constant term in the 
value of negativity.
In the case of mutual information \HartmanMIA\  the holographic result could
be used to predict a phase transition between the nonvanishing branch for $x\ge 1/2$
and zero for $x\le 1/2$.
For negativity we also expect a phase transition, but cannot fix the value of $x$ where
it would happen.
In any case, it would be interesting to have a holographic prescription for 
computing negativity\foot{
The behavior of the OPE coefficients at large central charge plays an important role
in reproducing the holographic entanglement result. Their exponential dependence on
the conformal dimension of the internal operator $h_p$ may lead to the difference between
the values of entanglement negativity in holographic theories and our result, which 
assumes that a certain conformal block gives a dominant contribution.
We thank   T. Faulkner and T. Hartman for discussions on this issue.}.

\bigskip
\bigskip
\noindent {\bf Acknowledgement:} We thank T. Faulkner, T. Hartman, C. Herzog, G. Moore, N. Nekrasov and E. Tonni for discussions
and correspondence. 
The work of G.P. and A. P. was supported in part by the Van Gogh France-Netherlands collaboration grant, project n. 28848PH, by the National Science Foundation under Grant No. PHY11-25915, and a vidi grant from NWO. The work of M.K. was partially supported by the ERC Advanced Grant "SyDuGraM", by IISN-Belgium (convention 4.4514.08) and by the ``Communaut\'e Fran\c{c}aise de Belgique" through the ARC program and by the National Science Foundation under Grant No. PHY11-25915.
A.P. and M.K. thank Simons Summer Workshop 2013, DAMTP Cambridge, KITP UCSB and CERN
for hospitality. 
A.P. thanks ENS Paris, IAS Princeton, Stony Brook University and the University of Chicago, and G.P. the University of Leiden, for hospitality during the completion of this work.

\appendix{A}{Investigating the asymptotic expansion of the conformal blocks at large $c$.}

\noindent There are two ways one may compute the conformal block at large $c$. One involves first expanding in $x$ and then taking the large central limit, making use of  \confblock.
Another way is to first take the large $c$ limit and then expand in $x$. The starting point in this case is the Heun equation \difeq. The latter approach was recently taken by the authors of  \LitvinovSXA\ and \MenottiKRA\  who attacked the problem in two distinct ways. As it turns out, whether one first takes the infinite central charge limit and then expand in the corss-ratio or vice versa is immaterial. The two limits appear to be commuting with each other and both methods yield the same result, \ie, the infinite $c$ limit of \confblock.

As explained in the Discussion section, we cannot directly apply \confblock\ to our case. The intermediate operator we are interested in corresponds to a null state at large $c$ and the quadratic correction in the cross-ratio $x$ diverges. One might expect that approaching the problem using the methods of \LitvinovSXA\ and \MenottiKRA\ adapted to our special case, would be possible to obtain 
a finite result. Here we see that this is not the case.

The authors of \LitvinovSXA\ showed that there is a one-to-one correspondence between confromal blocks at large central charge and the Painleve VI equation. To be precise they showed that the monodromy problem of the Heun equation can be mapped to the connection problem of the Painleve VI. Unfortunately, this approach cannot be succesfully applied here, because the Painleve solution for our case is given by a Taylor series expansion. This fact does not allow us to solve the connection problem order by order in $x$, as done in \LitvinovSXA\ for intermediate operators of generic dimension $h_p$.

We will thus use the method outlined in \MenottiKRA. Let us review here the basic ingredients of the approach. Suppose that we wish to determine $c_2$ such that the monodromy of the solutions of eq \difeq\ on a path encircling both $(0,x)$ is given by $\tr{M}=-2 \cos{\left[\Lambda_p\pi\right]}$. We can expand $T(z)$ for small $x$ as\foot{Here for simplicity we discuss the of small $x$. The case of small $y=1-x$ can be treated in an identical manner.}
\eqn\Tx{T(z)=T_0+T_1 x+ T_2 x^2+\OO(x^3)}
where
\eqn\defTi{\eqalign{T_0&\equiv {C_2(0)\over z^2(1-z)}\cr
T_1&\equiv {C_2(0)+zC_2'(0)\over z^3(1-z)}\cr
T_2&\equiv {2C_2(0)+2 zC_2'(0)+z^2 C_2''(0)\over 2 z^4(1-z)}\,.}}
Since we are interested in the limit $n\rightarrow 1$ we set $h_i=0$  
and for convenience defined $C_2(x)\equiv c_2(x)x(x-1)$ such that $C_2(0)={6\over c} h_p={1\over 4}\left(1-\Lambda_p^2\right)$.
We now consider the differential equation \difeq\ to leading order in $x$, namely,
\eqn\eqxzero{\psi''(z)+{C_2(0)\over z^2(1-z)} \psi(z)=0\,}
which for generic values of $C_2(0)$ is solvable by means of simple hypergeometric functions. 
The key observation of \MenottiKRA\ was that one can compute the monodromy matrix of the solutions
of eq. \eqxzero\ along a special path, depicted in Fig.2, for which $T(z)$ is not singular. With this choice of contour, the monodromy matrix is easily determined by considering the behavior of the solutions at infinity.
To leading order in $x$ one finds that
\eqn\trxzero{\tr{M^{(0)}}=-2\cos{\left[\Lambda_p\pi\right]}}
which implies that all the corrections in $\tr{M^{(0)}}$ should vanish.
These corrections can in principle be computed by solving eq.\eqxzero\ order by order in $x$.
Requiring that the $\tr{(\delta M)}$ vanishes at each order then determines $C_2(x)$ as a series expansion in $x$.
\vskip.0005in

\midinsert
{\epsfxsize1.8in\epsfbox{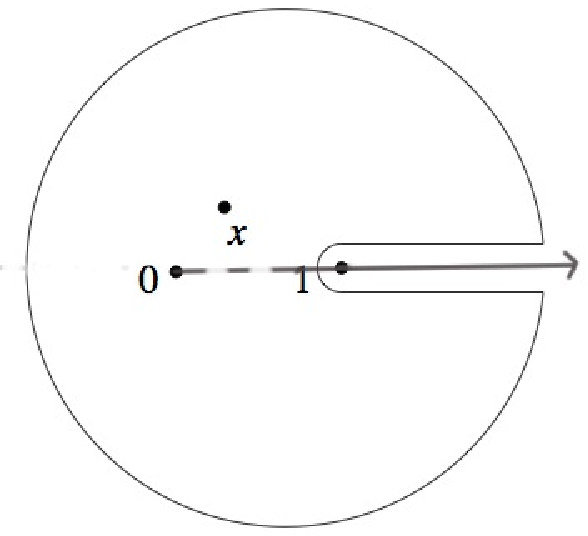}\vbox{This nice technique was developed in \MenottiKRA\ where $C_2(x)$ 

was determined up to quadratic order in $x$. The expres-

sions obtained there are in complete agreement with \LitvinovSXA. 

As remarked earlier however, one cannot directly apply 

the formulas in \refs{\LitvinovSXA-\MenottiKRA} when $\Lambda_p=2$. One reason this ca-

se is special can be traced to \eqxzero. 

%$C_2(0)=-{3\over 4}$ is a special point of 

%the hypergeometric equation where the standard set of solutions cannot be used. 
\vfill }
\hbox{\qquad\qquad \it Fig.2. The contour. }
\hfill 
}
\endinsert

\noindent
To be specific, when $C_2(0)=-{3\over 4}$ the differential equation 
\eqn\eqxzerob{\psi''(z)-{3\over 4 z^2 (1-z)} \psi(z)=0\,,}
can be placed in the standard hypergeometric form 
\eqn\sthyper{z(1-z)u''(z)+\left[c-(a+b+1)z\right]u'(z)-a b u(z)=0}
with $u(z)=z^{c\over 2} (1-z)^{1-c+a+b/2}\psi(z)$ and $(a,b,c)=({3\over 2},{1\over 2},3)$.
However, when $c$ is an integer, only one of the two independent solutions around $z=1$ is given by
a standard hypergeometric function. For the other solution a much more compicated series representation
exists. Here for simplicity we will use the following set of independent solutions
\eqn\sols{\eqalign{\psi_1(z)&={3\over 4} z^{3\over 2} (1-z)\, _2F_1\left({3\over 2},{5\over 2},2,1-z\right)\cr
\psi_2(z)&={3 i\pi^2\over 16} z^{3\over 2} \, _2F_1\left({3\over 2},{1\over 2},3,z\right)
}}
where the second solution is the standard solution around $z=0$. Their wronskian is  
\eqn\wab{w(\psi_1,\psi_2)\equiv \psi_1\psi_2'-\psi_1'\psi_2=-{3i\pi\over 4}\,.}
It is sometimes convenient to express the solutions in terms of the complete 
elliptic integrals of the first and second kind, \ie, 
\eqn\elldef{\eqalign{E[z]&=\int_0^{\pi\over 2} \sqrt{1-z \sin^2{\phi}} \cr
K[z]&=\int_0^{\pi\over 2}{d\phi\over \sqrt{1-z\,\sin^2{\phi}}}
}}
in the following way
\eqn\ellipticsol{\eqalign{\psi_1(z)&={1\over\pi} z^{-{1\over 2}}\left((z-2)E[1-z]+zK[1-z]\right)\cr
\psi_2(z)&=-i\pi z^{-{1\over 2}} \left((z-2)E[z]-2(z-1)K[z]\right)}\,.}
%The wronskian is equal to
%\eqn\wronskian{w_{12}\equiv \psi_1'(z)\psi_2(z)-\psi_1(z)\psi_2'(z)={3 i\over 2} \left(K[1-z]E[z]+E[1-z]K[z]-K[z]K[1-z]\right)}
The analytic continuation of $\psi_1(z)$ for $z>1$ above and below the axis can be simply written as
\eqn\anconsola{\psi_1^{\pm}(z)={3\over 4} z^{3\over 2}e^{\mp i \pi} (z-1)\,_2F_1\left({3\over 2},{5\over 2},2,1-z\right)\equiv -g_1(z)}
For the other solution we write
\eqn\anconsolb{\eqalign{\psi_2^\pm(z)=-i\pi z^{-{1\over 2}} \left((z-2)E[-e^{\mp i\pi}z]-2(z-1)K[-e^{\mp i\pi}z]\right)}}

To compute the monodromy matrix to leading order in $x$ we need the asymptotic behavior of the solutions at infinity.
To find the behavior for large $|z|>>1$ it is convenient to use the known asymptotic expansions
of the elliptic integrals and substitute $z$ by $z\rightarrow -z e^{\pm i\pi}$. The result is
\eqn\solinfty{\eqalign{\psi_1^{\pm}(z)&=-{z\over \pi}-{3\over4\pi} \ln{z}+{3\over 4\pi}(3-4\ln{2}) +\OO(z^{-1}\ln{z})\cr
\psi_2^{\pm}(z)&=\mp \pi z+\mp {3\pi\over 4} \ln{z}+\pm {3\pi\over 4}(3-4\ln{2})+{3 i\pi^2\over 4}+\OO(z^{-1}\ln{z})}\,.}
%where $s_i^{\pm}$ are the following constants
%\eqn\sidef{s_1^{\pm}={3\over 4\pi} \left(3-4\ln{2}\right),\qquad s_2^{\pm}=\pm\pi^2 s_1+{3\over 4}i\pi^2}
Denoting by $\Upsilon^{\pm}\equiv \pmatrix{\psi_1^\pm\cr \psi_2^\pm }$ the matrix of the two independent solutions, 
we can express the asymptotic behavior of the solution continued slightly {\it above} the real axis as
\eqn\upsilonp{\Upsilon^{+}_\infty\simeq B^{+} \Upsilon_0=\pmatrix{-{1\over\pi} &\quad 0\cr -\pi &\quad {3\over 4}i\pi^2} \pmatrix{z+{3\over 4} \ln{z}-{3\over 4}(3-4\ln{2})\cr 1}}
and that of the solution continued {\it below} the real axis as
\eqn\upsilonm{\Upsilon^{-}_\infty\simeq  B^{-}\Upsilon_0=\pmatrix{-{1\over\pi} &\quad 0 \cr \quad \pi & \quad  {3\over 4}i\pi^2 } \pmatrix{z+{3\over 4} \ln{z}-{3\over 4}(3-4\ln{2})\cr 1}
}  
We can now start from the solution defined at $z=\infty-i\epsilon$, eq.\upsilonm\ and take a full turn around infinity to go back to $z=\infty-i\epsilon$, but having now encircled both the origin and $x$ as can be seen in figure 1.
The turn around infinity yields
\eqn\upsilontildem{\tilde{\Upsilon}^{-}_\infty \simeq  \tilde{B}^{-}\Upsilon_0=\pmatrix{-{1\over\pi} &\quad {3\over 2} i \cr \quad \pi & \quad  -{3\over 4}i\pi^2 } \pmatrix{z+{3\over 4} \ln{z}-{3\over 4}(3-4\ln{2}) \ln{z}\cr 1}}
where the change in the constant terms is due to the change in the logarithm.
We now observe that 
\eqn\pmzerorel{\Upsilon^{+}=\left(B^{+}(\tilde{B}^{-})^{-1} \right) \,\tilde{\Upsilon}^{-}}
which allows us to compute the monodromy matrix to leading order in $x$,
\eqn\mona{M^{(0)}=\left(B^{+}(\tilde{B}^{-})^{-1} \right)=\pmatrix{-1 & -{2\over\pi^2}\cr \quad 0 &-1}} 
and confirm that $\tr{M^{(0)}}=-2$. 

Consider now the monodromy matrix $M$ of the solutions of eq. \difeq\ and its expansion in powers of $x$ 
\eqn\mtotal{M=M^{(0)}+x \delta M^{(1)}+x^2 \delta M^{(2)}+\OO(x^3)}
Since the desired trace $\tr{M}=\tr{M^{(0)}}=-2$ is achieved at leading order, it must be that $\tr{\delta M^{(1)}}=\tr{\delta M^{(2)}}=\cdots=0$.
The correction to the monodromy matrix at each order can be evaluated from the asymptotic behavior of the ``corrected'' solutions at infinity, \ie,
\eqn\cordeltam{\eqalign{\Upsilon^+_\infty+&x\delta\Upsilon^{(1),+}_\infty+x^2\delta\Upsilon^{(2),+}_\infty+\cdots=\cr
&=\left( M^{(0)}+x \delta M^{(1)}+x^2 \delta M^{(2)}+\cdots\right)
\left(\widetilde{\Upsilon}^-_\infty+x\delta\widetilde{\Upsilon}^{(1),-}_\infty+x^2\delta\widetilde{\Upsilon}^{(2),-}_\infty+\cdots\right)}}
Once $\delta M^{(i)}$ is expressed in terms of $C^{i}(0)$, solving $\delta M^{(i)}=0$ computes $C^{i}(0)$.

In what follows we focus on the first and second order correction to the monodromy matrix.
We start by considering equation \difeq\ to first order in $x$,
\eqn\eqxone{\psi''(z)+T_0(z)\psi(z)+x \, T_1(z)\psi(z)=0\,,}
where $T_0,\, T_1$ are given in \defTi. The solutions of \eqxone\ are of the form $\psi_i(z)+x \delta\psi_i(z)$ where $\psi_i$
are the solutions of the zeroth order in $x$ differential equation $\psi''+T_0\psi=0$ and $\delta\psi_i$ are the solutions of the inhomogeneous
differential equation
\eqn\deltapsieq{\delta\psi''+T_0\delta\psi=-T_1\psi}
%\eqn\solone{\psi_i(z)=\psi_i(z)+x \delta\psi_i(z)\,,}
which can be expressed as follows
\eqn\deltapsione{\delta\psi_i(z)={\psi_1(z)\over w_{12}} \int_1^z dz'\psi_2(z') T_1(z')\psi_i(z') -{\psi_2(z)\over w_{12}} \int_1^z dz' \psi_1(z')T_1(z')\psi_i(z')\,.}
Focusing on the asymptotic behavior of the solutions of \eqxone\ at infinity, we find that
%it is convenient to set $\delta\Upsilon^{\pm}=\pmatrix{\delta\psi_1^{\pm} \cr \delta\psi_2^{\pm}}$ and observe that
\eqn\deltaups{\delta\Upsilon^{\pm}_\infty= \Lambda_1^{\pm} \Upsilon^{\pm}_{\infty},\qquad \delta\widetilde{\Upsilon}^{-}_\infty=\Lambda_1^-\widetilde{\Upsilon}^-_\infty\,,}
where $\Lambda_1^\pm$ is the following matrix 
\eqn\defDelta{\Lambda_1^{\pm}=\pmatrix{\Delta^\pm_{12} & -\Delta^{\pm}_{11}\cr \Delta^\pm_{22}&-\Delta_{12}^\pm },\qquad \Delta_{ij}={1\over w_{12}} \int_1^\infty dz \psi_i(z) \, T_1(z) \psi_j(z)\,.}
%with
%\eqn\defDeltaij{\Delta_{ij}={1\over w_{12}} \int_1^\infty dz \psi_i(z) \, T_1(z) \psi_j(z)}
Substituting \deltaups\ into \cordeltam\ leads to 
\eqn\pmonerel{\delta M^{(1)}=\left(\delta Y^{+}_{\infty}-M^{(0)}\delta \tilde{Y}^{-}_\infty\right)=\Lambda_1^+ M^{(0)}-M^{(0)} \Lambda_1^-\,,}
while using \mona\ and taking the trace yields
\eqn\cona{\eqalign{\tr{\delta M^{(1)}}&=-{2\over\pi^2} \left(\Delta_{22}^+ -\Delta_{22}^- \right)=0
%\Rightarrow C_2'(0)={3\over 4}\,\, {I^{+}_{3,(22)}-I^{-}_{3,(22)}\over I^{+}_{2,(22)}-I^{-}_{2,(22)}}\,,
}}
With the help of eq. (A.51) we can express \cona\ 
%evaluate the integrals in \cona\ and rewrite it 
as follows
\eqn\conaa{\eqalign{&\left. U^{+}_{22}(f_0)-U_{22}^{-}(f_0)\right|_1^\infty=0
}}
where
\eqn\fonedef{f_0(v)={4\over 3}C_2'(0)(z-1)+\left(1-{8\over 3}C_2'(0)\right){z-1\over z}\,.}
and $U_{ij}$ is defined in (A.52).
Using the asymptotic expressions for the solutions $\psi^{\pm}$ at $z=1$ and $z=\infty$ given in Appendix A.3, the divergent terms cancel each other and we obtain
\eqn\conb{C_2'(0)={3\over 8}\,,} 
which matches the standard result for the conformal block at large $c$ for $\Lambda_p=-{3\over 4}$.

For the quadratic correction to the accessory parameter, we need to consider second order corrections
to the differential eq. \eqxzerob, namely,
\eqn\eqxtwo{\psi''(z)+T_0(z)\psi(z)+x T_1(z)\psi(z)+x^2 T_2(z)\psi(z)=0}
 with $T_0,\, T_1,\,T_2$ given in \defTi. The solutions of \eqxtwo\ are
\eqn\soltwo{\psi_i^{(2)}(z)=\psi_i(z)+x \delta\psi_i^{(1)}(z)+x^2\delta\psi_i^{(2)}(z)}
with $\delta\psi^{(1)}_i$ as in \deltapsione\ and $\delta\psi^{(2)}_i$ equal to
\eqn\deltapsitwo{\delta\psi_i^{(2)}={\psi_1\over w_{12}} \left(\int_1^z \psi_2 T_2\psi_i+ \int_1^z \psi_2 T_1 \delta\psi_i^{(1)}\right)-{\psi_2\over w_{12}} \left(\int_1^z \psi_1 T_2\psi_i+ \int_1^z \psi_1 T_1 \delta\psi_i^{(1)}\right)\,.}
To find the asymptotic behavior of the solutions at infinity note that
\eqn\deltaupsb{\delta\Upsilon^{(2)\pm}_\infty=\Lambda_2^\pm \Upsilon^\pm_\infty,\qquad \delta\widetilde\Upsilon^{(2)-}_\infty=\Lambda_2^{-}\widetilde{\Upsilon}_\infty^-}
where
\eqn\lambdaded{\Lambda_2^\pm=Z^\pm+\Theta^\pm,\qquad
Z^\pm=\pmatrix{Z_{12}^\pm & -Z_{11}^\pm \cr
Z^\pm_{22} & -Z_{12}^\pm} ,\qquad \Theta^\pm=\pmatrix{\Theta_{12}^\pm & -\Theta_{11}^\pm\cr
\Theta_{22}^\pm &-\Theta_{21}^\pm}
}
and
\eqn\ThetaZdef{Z_{ij}^\pm={1\over w_{12}} \int_1^\infty \psi_i^\pm T_2\psi_j^\pm, \qquad\quad\Theta_{ij}^\pm={1\over w_{12}}\int \delta\psi^{(1)\pm}_i\,T_1\psi_j^\pm}
Using eqs. \deltaupsb, \ThetaZdef\ together with the relations $\Delta_{11}^+=\Delta_{11}^-$ (which is due to \anconsola) and $\Delta_{22}^+=\Delta_{22}^-$ 
(which ensures that $\delta M^{(1)}=0$), results in
\eqn\conb{\eqalign{\tr{\delta M^{(2)}}&={2\over\pi^2}\left[(\Delta_{12}^+-\Delta_{12}^-)(\Delta_{22}^-+\pi^2\Delta_{12}^-)+\right.\cr
&\left.+{1\over 2}\left((\Theta_{12}^--\Theta_{21}^-)-(\Theta_{12}^+-\Theta_{21}^+)\right)+
\left((\Theta_{22}^-+Z_{22}^-)-(\Theta_{22}^++Z_{22}^+)\right)\right]}}
With the help of Appendix A we can express the different terms appearing in \conb\ as follows
\eqn\simconbt{\eqalign{w_{12}(\Delta_{12}^+-\Delta_{12}^-)&(\Delta_{22}^-+\pi^2\Delta_{12}^-)=\left.(U_{12}^+(f_0)-U_{12}^-(f_0))(U_{22}^-(f_0)+\pi^2 U_{12}^-(f_0))\right|_1^\infty\cr
w_{12}(\Theta_{12}^{\pm}-\Theta_{21}^{\pm})&= \left.S_{12}^{\pm}(f_0,g_0)-S_{21}^{\pm}(f_0,g_0)\right|_1^\infty\cr
w_{12}\left(\Theta_{22}^\pm+Z_{22}^\pm\right)&=\left. S_{22}^\pm(f_0,g_0)+{9\over 16} \left(S_{22}^\pm(f_4,g_4)-S_{22}^\pm(\hat{f}_4,\hat{g}_4)-\SS_{22}^\pm\right)-C_2''(0){1\over 2} U_{22}^{\pm}(f_2)\right|_1^\infty
}}
where $U_{ij}(f),S_{ij}(f,g),\SS_{ij}$ are defined in (A.52), (A.61) and (A.73) whereas
\eqn\fidef{\eqalign{f_0&={1\over 2}(z-1),\qquad g_0={1\over 4}(z-1)-{1\over 2}{z-1\over z},\qquad f_2=-{4\over 3}z+4-{8\over 3z}\cr
f_4&={4\over 3}(1-{1\over z}),\qquad g_4={16\over 9}-{8\over 9}{1\over z^3}+{4\over 9}{1\over z^2}-{4\over 3}{1\over z},\cr
\hf_4&={z\over 3}+{1\over 3}-{2\over 3 z},\qquad \hg_4={16\over 9}-{13\over 9 z}-{1\over 9 z^2}-{2\over 9 z^3}
}}
Finally, using the asymptotic expressions for the solutions $\psi^{\pm}_i$ as well as for  $p^\pm_i,q^\pm_i$ at $z=1$ and $z=\infty$ given in Appendix A.3 we find that all divergent terms cancel out except for the ones coming from the first order terms, \ie, $(\Delta_{12}^+-\Delta_{12}^-)(\Delta_{22}^-+\pi^2\Delta_{12}^-)$ (see (A.82)).
Hence the divergence at quadratic order persists, hinting at the non-analyticity of the accessory parameter $C_2(x)$.

\subsec{Analytic computation of integrals with hypergeometric functions.}

\noindent The following integrals are necessary for computing the first and second order corrections to the accessory parameter $C_2(x)$
\eqn\Itwothreedef{I^\pm_{m,ij}(z)=\int_1^z dz {\psi_i^{\pm}(z) \psi_j^{\pm} (z)\over z^m \,(z-1)},\qquad m=2,3,4\,,} 
and
\eqn\Idelta{\delta I_{ij}^{\pm}(z)=\int_1^z \delta\psi^{(1),\pm}_i T_1 \psi_j^{\pm},\qquad\quad T_1=-{3\over 8}{1\over z^2(z-1)}+{3\over 4}{1\over z^3(z-1)}\,.}
A method for computing such integrals was developed in \MenottiKRA. Here we will use and expand this method as is necessary to treat the special case $\Lambda_p=2$. We should mention here that such integrals are in general divergent. We will see that in most divergences cancel when evaluating the corrections to the accessory parameter.

Consider a coordinate transformation from the $z$--variable to some new variable $v$ which also depends on some arbitrary parameter $a$ such that
\eqn\zv{z\equiv z(v,a)=v+f(v) a+g(v)a^2+\OO(a^3)\,.}
The zeroth order differential equation $\psi''(z)+T_0(z)\psi(z)=0$ under this change of variable becomes
\eqn\diftr{\ddot{t}(v,a)+R(v,a) t(v,a)=0\,,}
where the dots denote differentiation with respect to $v$ and
\eqn\defRt{\eqalign{R(v,a)&=\left({\p z\over \p v}\right)^2 T_0(z(v,a)) +{1\over 2} \{z,\,v\},\cr
 t_i(v,a)&=\left({\p z(v,a)\over \p v}\right)^{-{1\over 2}}\psi_i(z(v,a))\,.}}
Here $\{z,v\}$ is the Schwartzian of the transformation defined as follows
\eqn\schw{\{z,v\}\equiv {{\ddot z}\over {\dot z}} -{3\over 2} \,\left( {\ddot{z}}\over \dot{z}\right)^2\,.}
From the definition and properties of the transformation \zv\ and \defRt\ it follows that
\eqn\Ra{R(v,a)=T_0(v)+a R_1(v)+a^2 R_2(v)+\OO(v^3)}
with
\eqn\ronetwodef{\eqalign{R_1(v)&={(6-9v)f(v)+2v(v-1)\left(3 f'(v)+v^2(v-1)f'''(v)\right)\over 4 v^3(v-1)^2}\cr
R_2(v)&={(6-9v)g(v)+2v(v-1)\left(3 g'(v)+v^2(v-1)g'''(v)\right)\over 4 v^3(v-1)^2}+\cr
&+{3(3-8v+6 v^2)f(v)^2-6v(2-5v+3 v^2)f'(v)f(v)\over 4 v^4(v-1)^3 }+\cr
&+{v^2(v-1)^2\left(3 f'(v)^2-3v^2(v-1)f''(v)^2-2(v-1)v^2f'(v)f'''(v)\right)\over 4 v^4(v-1)^3}
}}
Note also that $\lim_{a\rightarrow 0} t(v,a)=\psi(v)$. 

Let us now differentiate \diftr\ with respect to $a$ and take the limit $a\rightarrow 0$ to obtain
\eqn\difnew{{\ddot{\phi}}_i+T_0 {\dot{\phi}}_i+R_1 \psi_i=0,}
where we set
\eqn\phidef{\phi_i(v)\equiv \left. {\p t_i(v,a)\over\p a}\right|_{a=0}\,,}
and used $t_i(v,a=0)=\psi_i(v)$.
Multiplying \difnew\ with $\psi_j$ and integrating by parts leads to\foot{We assume that ${\p\over\p a}\,\left.{\p\over\p v}\right|_{a=0}={\p\over\p v} \,\left.{\p\over\p a}\right|_{a=0}$} 
\eqn\intRon{\int_1^z \psi_j R_1 \psi_i=\left.\psi_j'(z)\phi_i(z)-\psi_j \phi'(z)\right|_1^z}
%&=\left.f(z)\left(\psi'_i\psi'_j-\psi_i\psi_j''\right)-{1\over 2} f'(z)\left(\psi'_i\psi_j+\psi_i\psi_j'\right)+{1\over 2} f''(z)\psi_i\psi_j\right|%_1^z}}
Further using \zv\ and \defRt\ to evaluate $\phi_i$ yields
\eqn\intRone{\int_1^z \psi_j R_1 \psi_i=\left.U_{ij}(f)\right|_1^z}
where we defined
\eqn\defUij{U_{ij}(f)=f(z)\left(\psi'_i\psi'_j-\psi_i\psi_j''\right)-{1\over 2} f'(z)\left(\psi'_i\psi_j+\psi_i\psi_j'\right)+{1\over 2} f''(z)\psi_i\psi_j}
%where to obtain the last equality we used both \zv\ and \defRt. 
To confirm that $U_{ij}(f)$ is symmetric under the exchange 
$\psi_i\rightarrow \psi_j$, recall that $\psi_i\psi_j''=\psi_i''\psi_j$ for two solutions $\psi_{i,j}$ with constant wronskian \wab.

Computing integrals of the type \Itwothreedef\ is now a straightforward exercise; one simply needs to choose $R_1(v)$, or rather $f(v)$, appropriately and employ \intRone. Thus the original problem has been mapped to the problem of finding a particular solution to the inhomogeneous differential equation which determines $f(v)$ for a specific choice of the function $R_1(v)$, \ie, \Ra.

Finding a suitable $f(v)$ for the integrals in \Itwothreedef\ with $m=2,3$ is relatively easy. The choices $f(v)=-{4\over 3}v+4-{8\over 3 v}$ and $f(v)={4\over 3}(1-{1\over v})$ respectively yield
\eqn\Rtwothree{\eqalign{R_1&={1\over v^2(v-1)},\qquad\quad f(v)=f_2(v)\equiv-{4\over 3}v+4-{8\over 3 v}\cr
R_1&={1\over v^3(v-1)},\qquad\quad f(v)=f_3(v)\equiv {4\over 3}(1-{1\over v})\,.
}}
%Using \Rtwothree\ in combination with \intRone\ we can compute two linearly independent combinations of $I_{2,(ij)},\, I_{3,(ij)}$ and %from them $I_{2,(ij)},\,I_{3,(ij)}$ separately\foot{These two choices of $f(v)$ have been exploited in \MenottiKRA.\ in a different way. A %degeneracy which occurs for the special case $\Lambda_p=2$... }. 
It is also possible to directly compute the necessary linear combination of integrals appearing in $\Delta_{ij}$ (defined in \defDelta) by choosing $f(v)=f_0(v)\equiv {4 C_2'(0)\over 3}(v-1)+{3-8 C_2'(0)\over 3}{v-1\over v} $ such that $R_1(v)=T_1(v)$.

The case $m=4$ in \Itwothreedef\ is special because a simple solution for $f(v)$ cannot be easily found.  We will describe a slightly different technique for evaluating $I_4$ in the following subsection. We conclude here by explaining how to compute integrals like the one in \Idelta. 

The starting point is again eq.\diftr\ which we now differentiate twice with respect to $a$ to obtain
\eqn\difnewb{\ddot{\chi}_i+T_0\chi_i+2 R_1 \phi_i +2 R_2 \psi_i=0\,.}
In writing \difnewb\ we used the following definitions
\eqn\chidef{\chi(v)\equiv\left. {\p^2 t(v,a)\over \p a^2}\right|_{a=0},\qquad \phi(v)\equiv\left. {\p t(v,a)\over \p a}\right|_{a=0},\quad \left. t(v,a)\right|_{a=0}=\psi(v)\,.}
Let us note here that $\phi$ satisfies a differential equation similar to the one $\delta\psi$ satisfies.
In fact, when $R_1=T_1$, \difnew\ coincides with \deltapsieq\ and as a result $\phi_i$ can be identified with $\delta\psi_i$.
It is now clear how \difnewb\ can help us compute \Idelta. We first choose $f(v)$ in such a way that $R_1=T_1$ and $\phi=\delta\psi_i^{(1)}$. As previously mentioned, this can be achieved by 
\eqn\fronetone{f(v)=f_0(v)={4 C_2'(0)\over 3}(v-1)+{3-8 C_2'(0)\over 3}{v-1\over v} ,\qquad\Rightarrow_{C_2'(0)={3\over 8}}\qquad f_0(v)={1\over 2}(v-1)\,.}
We then multiply \difnewb\ with $\psi_j$ and integrate by parts to obtain
\eqn\intRonephi{\int_1^z \delta\psi_i^{(1)}T_1\psi_j={1\over 2}\left.\left( \psi'_j\chi_i-\psi_j\chi'_i \right)\right|_1^z-\int_1^z \psi_i R_2\psi_j\,.}
As long as we can evaluate $\int_1^z \psi_i R_2\psi_j$, we can also evaluate \intRonephi. Note however that $R_2$ can take any form we like since  \ronetwodef, as a linear differential equation for $g(v)$, admits a solution for any inhomogeneous term, at least in principle. Here we choose
\eqn\grtwo{g(v)=g_0(v)\equiv {1\over 4} (v-1)-{1\over 2}{v-1\over v}\,,} 
which results in
\eqn\rtwochoice{R_2={9\over 16}{1\over v^4(v-1)}-{3\over 8}{1\over v^3(v-1)}\,.}
With the choice \grtwo\ the integral on the right hand side of \intRonephi\ is of the type of \Itwothreedef\ with $m=3,4$, which
we know how to evaluate.
Finally, combining \rtwochoice\ together with \fronetone\ and \intRonephi\ while taking into account \zv\ and \defRt\ leads to
\eqn\intRonephif{\int_1^z \delta\psi_i^{(1)}T_1\psi_j=S_{ij}(f_0,g_0)-{9\over 16}I_{4(ij)}+{3\over 8}I_{3(ij)}}
where $S_{ij}(f,g)$ is defined as
\eqn\sijdef{\eqalign{S_{ij}(f,g)&\equiv \left. {1\over 2}\psi_j'\left[ \psi_i ({3\over 4} f'^2- g')+\psi_i'(2g-ff')+ f^2\psi_i'' \right]\right|_1^z-\cr
&-{1\over 2}\left.\psi_j \left[\psi_i({3\over 2} f' f''- g'')+\psi_i'(g'-{1\over 4}f'^2- ff'')+\psi_i''(2g+ f f')+ f^2\psi_i'''\right] \right|_1^z
}}

\subsec{Computing the integral $I_4$.}

\noindent Consider the following integral
\eqn\intthreepp{I_{3}(\phi,\psi)\equiv \int_1^z dv\,{\phi_i(v) \psi_j(v)\over v^3(v-1)}\,,}
where $\psi_i$ is a solution of \eqxzerob and $\phi_i$ satisfies the inhomogeneous differential equation
\eqn\deltapsione{\phi_i''+T_0\phi_i+V_1\psi_i=0\,.} 
with $V_1={1\over z^3(z-1)}$. 

We explained how to evaluate integrals of this form at the beginning of Appendix A.
We start from the original differential equation \eqxzerob\ and perform a change variable as in \zv,
choosing $f(v)$ such that $R_1$ in \Ra\ coincides with $V_1$. We then pick $g(v)$ such
that the  integral $\int_1^z \psi_i\psi_j R_2$ is expressed in terms of the known integrals $I_2,I_3$ and/or the sought for integral $I_4$. In practice we find that a convenient choice is
\eqn\fgphi{f(v)={4\over 3}(1-{1\over v})\equiv f_4(v),\qquad g(v)={16\over 9}-{8\over 9}{1\over v^3}+{4\over 9}{1\over v^2}-{4\over 3}{1\over v}\equiv g_4(v)}
and leads to 
\eqn\intphione{\int_1^z {\phi_i\psi_j\over v^3(v-1)}=S_{ij}(f_4,g_4)-{7\over 3} I_{4}}
where $S_{ij}(f,g)$ is defined in \sijdef.

Consider now an alternative, slightly more involved method to evaluate the same integral, \intthreepp.
We start from \eqxzerob\ but instead of performing a change of variable, we induce a change in $T_0$ parametrized by $a$, \ie,
\eqn\eqxzeroba{u''+{3\over 4 z^2(z-1)} (1+a s_0) u=0,\,}
with $s_0$ an arbitrary number. The two independent solutions of \eqxzeroba\ can be expressed in terms 
of hypergeometric functions as follows
\eqn\soleqxzeroba{\eqalign{u_1(a,z)&={3\over 4}(z-1)z^{1+\sqrt{4+3 a s_0}\over 2} \,_2F_1\left[{1+\sqrt{4+3 a s_0}\over 2},{3+\sqrt{4+3 a s_0}\over 2},2,1-z\right]\cr
u_2(a,z)&={3\over 16} i \pi^2 z^{1+\sqrt{4+3 a s_0}\over 2}\, _2F_1\left[{-1+\sqrt{4+3 a s_0}\over 2},{1+\sqrt{4+3 a s_0}\over 2},1+\sqrt{4+3 a s_0},z\right]\,. 
}}
The basis of the solutions is selected so that it reduces to \sols\ when $s_0=0$. Taylor expansion around the point $a=0$ yields $u(a,z)=\psi(z)+a p(z)+a^2 q(z)+\OO(a^2)$, where the functions $p(z),q(z),\cdots$ can be computed explicitly.

Next we change variables from $z$ to $v$ according to \zv\ with the same parameter $a$ appearing in \eqxzeroba. We are interested in producing an additional relation between the integrals $I_4$ and $I_3(\phi,\psi)$, we thus choose the pair $(f,g)$ such that $R_1$ in \Ra\ is {\it identical} to $V_1$ and $R_2$ is proportional to ${1\over v^4(v-1)}$. It is straightforward to verify that the following set of functions $(\hf_4,\hg_4)$ satisfy the above requirements, \ie,
\eqn\fgtwo{\eqalign{f(v)&=s_0 v+{1\over 3}(4-9 s_0)+{2\over 3}(-2+3 s_0){1\over v}\equiv \hf_4(v),\cr
g(v)&=\left({16\over 9}+{4\over 3} s_0- 4s_0^2\right)-\left({4\over 3}+{4\over 3}s_0-3 s_0^2\right){1\over v}+\left({4\over 9}-{8\over 3} s_0+3 s_0^2\right){1\over v^2}-\cr
&\qquad\qquad\qquad -\left({8\over 9}-{8\over 3} s_0+2 s_0^2\right){1\over v^3}\equiv \hg_4(v)\,.
} }
Following the approach of Appendix A we differentiate \eqxzeroba\ twice with respect to $a$ and evaluate at $a=0$, which yields
\eqn\eqxzerobatwo{\ddot{\chi_i}+T_0\chi_i+2 R_1\phi_i+2 R_2\psi_i=0\,,}
with
\eqn\ronertwochi{\chi_i\equiv \left.{\p^2 t_i\over \p a^2}\right|_{a=0},\phi_i\equiv \left.{\p t_i\over \p a}\right|_{a=0},\quad
R_1={1\over v^3(v-1)},\quad R_2={7-9s_0^2\over 3 v^4(v-1)}\,.}
Once more, $R_1$ being equal to $V_1$ implies that $\phi_i$ in \eqxzerobatwo\ is a solution of \deltapsione.
Muptiplying with $\psi_j$ and integrating leads to  
\eqn\intphitwo{\eqalign{\int_1^z {\phi_i\psi_j\over v^3(v-1)}&={1\over 2}\left.\left(\dot{\psi}_j\chi_i-\psi_j\dot{\chi}_i\right)\right|_1^z-{1\over 3}(7-9 s_0^2)I_{4,ij}=.\cr
&={1\over 2}\dot{\psi}_j\left(2 q_i-p_i \dot{f}_2+2\dot{p}_i f_2\right)-{1\over 2}\psi_j\left(2\dot{q}_i+\dot{p}\dot{f}_2-p_i\ddot{f}_2+2\ddot{p}_i f_2\right)+\cr
&+S_{ij}(f_2,g_2)-{1\over 3}(7-9 s_0^2)I_{4,ij}
}}
$I_4$ is then readily computed using the linearly independent equations \intphitwo\ and \intphione,
\eqn\ifoureq{I_{4,(ij)}={S_{ij}(f_4,g_4)-S_{ij}(\hf_4,\hg_4)-\SS_{ij}\over 3s_0^2}\,}
where we defined
\eqn\SSij{\SS_{ij}\equiv {1\over 2}\dot{\psi}_j\left(2 q_i-p_i \dot{f}_2+2\dot{p}_i f_2\right)-{1\over 2}\psi_j\left(2\dot{q}_i+\dot{p}\dot{f}_2-p_i\ddot{f}_2+2\ddot{p}_i f_2\right)} 
Thus far, we kept $s_0$ arbitrary. In the actual computation, we chose $s_0={1\over 3}$.

We have finally shown how to compute all the integrals necessary for the evaluation of the quadratic corrections to the accessory parameter $C_2$. The rest is a matter of bookkeeping.

\subsec{List of asymptotic expressions.}

\item 1. Asymptotics of the solutions
\eqn\solone{\eqalign{\psi_1^{\pm}(z)&\simeq_{z\sim\infty}-{z\over \pi}-{3\over4\pi} \ln{z}+{3\over 4\pi} \left(3-4\ln{2}\right)+{9\over 32\pi}{\ln{z}\over z}-{3\over 64\pi}\left(1-24 \ln{2}\right){1\over z}+\cdots \cr
\psi_1^{\pm}(z)&\simeq_{z\sim 1} -{3\over 4}(z-1)+{9\over 32}(z-1)^2+\cdots
}}
\eqn\soltwo{\eqalign{\psi_2^{\pm}(z)&\simeq_{z\sim\infty}\mp \pi z+\mp {3\pi\over 4} \ln{z}\pm{3\pi\over 4} \left(3-4\ln{2}\right)+{3\over 4}i\pi^2 \cr
&\quad\qquad\pm {9\pi\over 32} {\ln{z}\over z}+\left(\mp {3\pi\over 64}\left(1-24 \ln{2}\right)-{9i\pi^2\over 32}\right){1\over z}+\cdots \cr
\psi_2^{\pm}&\simeq_{z\sim 1} i \pi-{3i\pi\over 4}\ln{[z-1]}(z-1)+\left({i\pi\over 4}(-5+12\ln2) \mp {3\pi^2\over 4}\right)(z-1)+\cdots
}}

\item 2. Asymptotics of $p_2^{\pm}(z)$.
\eqn\ptwoas{\eqalign{ p_2^{\pm}&\simeq_{z\sim\infty} p_{21}^{\pm} z+p_{22}^{\pm}\ln{z}+p_{23}^{\pm}+\cdots\cr
p_2^{\pm}&\simeq_{z\sim 1} P^{\pm}_{21}+P_{22}^{\pm} (z-1)\ln{[z-1]}+P_{23}^{\pm}(z-1)+\cdots
}}
where
\eqn\pdef{\eqalign{p_{21}^\pm&=\pm {\pi\over 24}(5-12\ln{2})-{i\pi^2\over 8}\cr
 p_{22}^\pm &=\pm{3\pi\over 32}(-1-8\ln{2})-{3i\pi^2\over 32}\cr
p_{23}^\pm &=\pm {3\pi\over 32} \left(-5-8\ln{2}(-1+2\ln{2})\right)+{3i\pi^2\over 8}
}}
and
\eqn\Pdef{\eqalign{P_{21}^\pm&={i\pi\over 24}(-5+12\ln{2})\cr
 P_{22}^\pm &=-{3i\pi\over 32}(1+4\ln{2})\cr
P_{23}^\pm &=\pm{3\pi^2\over 32}(-1-4\ln{2})-{i\pi\over 96}\left(-25+9\pi^2+24(1-6\ln{2})\ln{2}\right)
}}

\item 3. Asymptotics of $q_2^\pm(z)$.
\eqn\qtwoas{\eqalign{ q_2^{\pm}&\simeq_{z\sim\infty} q_{21}^{\pm} z+q_{22}^{\pm}\ln{z}+q_{23}^{\pm}+\cdots\cr
q_2^{\pm}&\simeq_{z\sim 1} Q^{\pm}_{21}+Q_{22}^{\pm} (z-1)\ln{[z-1]}+Q_{23}^{\pm}(z-1)+\cdots
}}
where
\eqn\qdef{\eqalign{q_{21}^\pm&=\pm {\pi\over 1152}\left(-71+12\pi^2+156\ln{2}-144\ln^2{2}\right)-{i\pi^2\over 384}(-13+24\ln{2})\cr
q_{22}^\pm &=\pm{\pi\over 512}\left(4\pi^2-3(-1+8\ln{2}+64\ln^2{2})\right)-{3i\pi^2\over 512}(1+8\ln{2})\cr
q_{23}^\pm &=\pm {\pi\over 512} \left(63-48\ln{2}(3-2\ln{2}+4\ln^2{2})+4\pi^2(-3+4\ln{2})-84 \zeta(3)\right)+\cr
&+{i\pi^2\over 256}\left(\pi^2-3(7-12\ln{2}+8\ln^2{2})\right)
}}
and
\eqn\Qdef{\eqalign{Q_{21}^\pm&=-{i\pi\over 1152} \left(-71+3\pi^2+12(13-12\ln{2})\ln{2}\right)\cr
Q_{22}^\pm &={i\pi\over 512}(3+\pi^2-12\ln{2}(1+4\ln{2}))\cr
Q_{23}^\pm &=\pm{\pi^2\over 512}\left(\pi^2-3(-1+4\ln{2}+16\ln^2{2})\right)-\cr
&-{i\pi\over 4608}\left(283-1728\ln^3{2}+6\pi^2(2+42\ln{2})+96\ln{2}(-7+3\ln{2})-756\zeta(3)\right)
}}
where $\zeta(3)$ represents the Riemann $\zeta$-function.
\item 4. Additional Relations
\eqn\extrasb{\eqalign{
w_{12}(\Theta_{12}^--\Theta_{21}^-)&=w_{12}(\Theta_{12}^+-\Theta_{21}^+)=0\cr
S_{22}^-(f_0,g_0)-&S_{22}^+(f_0,g_0)=-{3i\pi^3\over 4},\qquad U_{22}^-(f_2)-U_{22}^+(f_2)=4i\pi^3\cr
S_{22}^-(f_4,g_4)-&S_{22}^+(f_4,g_4)={10 i\pi^3\over 3},\qquad S_{22}^-(\hf_4,\hg_4)-S_{22}^+(\hf_4,\hg_4)={17 i\pi^3\over 6}\cr
\SS_{22}^--\SS_{22}^+&={i\pi^3\over 12}\left(-7+12\ln{2}+{9\pi^2\over 32}\right)\,.
}}
Finally, we see that divergent terms remain in the following expression
\eqn\extrasa{\eqalign{&w_{12}(\Delta_{12}^+-\Delta_{12}^-)(\Delta_{22}^-+\pi^2\Delta_{12}^-)=\cr
&=\left({1\over 4}(11-12\ln{2})-{3\over 4}\lim_{z\rightarrow\infty}\ln{z}\right)\times {\pi^2\over 8} \left((11-12 \ln{2})+3\lim_{z\rightarrow 1}\ln{(z-1)}\right)\,.
}}

\footatend\vfill\supereject\immediate\closeout\rfile\writestoppt
\baselineskip=14pt\centerline{{\bf References}}\bigskip{\frenchspacing%
\parindent=20pt\escapechar=` \input refs.tmp\vfill\eject}\nonfrenchspacing
\bye